\documentclass[12pt,preprint]{aastex}

\begin{document}

\title{Vorticity of IGM Velocity Field on Large Scales}

\author{Weishan Zhu$^{1,2}$, Long-long Feng$^{1,3}$ and Li-Zhi Fang$^{2}$}

\altaffiltext{1}{Purple Mountain Observatory, Nanjing, 210008, China}
\altaffiltext{2}{Department of Physics, University of Arizona, Tucson, AZ 85721}
\altaffiltext{3}{Department of Physics,Texas A\&M University,College Station,TX 77843}

\begin{abstract}

We investigate the vorticity of the IGM velocity field on large
scales with cosmological hydrodynamic simulation of the concordance
model of $\Lambda$CDM. We show that the vorticity field is
significantly increasing with time as it can effectively be
generated by shocks and complex structures in the IGM. Therefore,
the vorticity field is an effective tool to reveal the nonlinear
behavior of the IGM, especially the formation and evolution of
turbulence in the IGM.  We find that the vorticity field does not
follow the filaments and sheets structures of underlying dark matter
density field and shows highly non-Gaussian and intermittent
features. The power spectrum of the vorticity field is then used to
measure the development of turbulence in Fourier space. We show that
the relation between the power spectra of vorticity and velocity fields
is perfectly in agreement with the prediction of a fully developed
homogeneous and isotropic turbulence in the scale range from $0.2$
to about $3 h^{-1}$ Mpc at $z\sim 0$. This indicates that cosmic
baryonic field is in the state of fully developed turbulence on
scales less than about 3 $h^{-1}$ Mpc. The random field of the
turbulent fluid yields turbulent pressure to prevent the
gravitational collapsing of the IGM. The vorticity and turbulent
pressure are strong inside and even outside of high density regions.
In IGM regions with 10 times mean overdensity, the turbulent
pressure can be on an average equivalent to the thermal
pressure of the baryonic gas with a temperature of $1.0\times 10^5$
K. Thus, the fully developed turbulence would prevent the baryons in 
the IGM from falling into the gravitational well of dark matter halos. 
Moreover, turbulent pressure essentially is dynamical and non-thermal, which
makes it different from pre-heating mechanism as it does not affect
the thermal state and ionizing process of hydrogen in the IGM.

\end{abstract}

\keywords{cosmology: theory - intergalactic medium - large-scale
structure of the universe - methods: numerical}

\section{Introduction}

Gravity is curl-free in nature, therefore it is unable to trigger
vorticity within the velocity field of a cosmic flow. On the linear
order of cosmological perturbation theory, the vorticity will
inevitably decay due to the expansion of the universe. The linear
velocity fields of the cosmic flow should be irrotational. In the
nonlinear regime of clustering, vorticity can be generated in the
collisionless dark matter field when multi-streaming occurs at shell
crossing (Binney, 1974, Pichon \& Bernardeau, 1999). However, there
is no way to directly map the vorticity of the dark matter field to
the baryon field, most of which is the intergalactic medium (IGM).

In the context of fluid dynamics, vorticity can be generated if the
gradient of the mass density and the pressure gradient of cosmic
flow are not aligned (Landau \& Lifshitz 1987). Namely, vorticity
results from the complex structures of fluid flow like curved
shocks. Recently, it has been revealed that in the nonlinear
regime, the cosmic baryon fluid at low redshift does contain such
complex structures (e.g. He et al 2004). Therefore, one expects that
vorticity would be present and evolve extensively in the cosmic
baryonic field. The vorticity of the intracluster medium (ICM)
has been studied in topics related to possible mechanism of
generating magnetic field of galaxies or clusters (Davis \& Widrow
2000; Ryu, et al 2008). Although these works show that the vorticity can form
in the ICM, the formation and evolution of vorticity in the IGM
is still unknown.

In addition, no studies have been done on the relation between
vorticity and turbulence in a cosmic baryon fluid. Actually, vortices
generally are considered a fundamental ingredient of turbulence
and the fluctuations of the vorticity field is an important indicator to
describe the turbulence of fluid (e.g. Batchelor, 1959, Schmidt 2007).
On the other hand, the study of the turbulence of cosmic fluid on
large scales has seen a lot of progress in recent years. The
fluctuations of the velocity field of the baryon fluid beyond the
Jeans length is shown to be extremely well described by the
She-Leveque (SL) scaling (He, et al 2006), which is the generalized
scaling of the classical Kolmogorov's 5/3-law of fully developed
turbulence (She \& Leveque 1994). The non-Gaussian features of the
density field in baryon flows are found to be in good agreement with
the log-Poisson cascade (Liu \& Fang 2008), which characterizes
statistically the hierarchical structure in fully developed
turbulence (Dubrulle 1994; She \& Waymire 1995; Benzi et al. 1996).
Observationally, the intermittence of Ly$\alpha$ transmitted flux of
QSO absorption spectrum can also be well explained in terms of
log-Poisson hierarchy cascade(Lu et al. 2009). These results suggest
that the dynamical behavior of the IGM is similar to a fully
developed turbulence in inertial ranges. Therefore, it would be
worthwhile to investigate the vorticity fields of the turbulent
cosmic fluid.

An important problem related to the vorticity fields of a
turbulent fluid is the impact of the turbulent pressure on the
clustering of cosmic fluid. It is well known that the random
velocity field of a turbulent fluid will play a similar role as
thermal pressure and prevent the gravitational collapse in such a 
fluid (Chandrasekhar, 1951; Bonazzola et al. 1992). In the ICM, this
effect has been studied with hydrodynamic simulations (Dolag et al
2005; Iapichino \& Niemeyer 2008; Cassano 2009, and reference
therein). However, in these works the turbulent pressure is
directly identified with the RMS baryon velocity. This
identification may be reasonable for the ICM; however, it would be
a poor relation on scales larger than clusters, as the RMS baryon velocity
cannot separate the velocity fluctuations due to bulk motion from
that of turbulence. Obviously, the bulk motion is not going to
prevent gravitational collapsing. Since the dynamical equation of
vorticity is free from gravity, the vorticity field provides an
effective method to pick up the velocity fluctuations within a
turbulent flow. The power spectrum of the vorticity field yields a
measurement on the scale of velocity fluctuations where
turbulence is fully developed. Using this method,
we can estimate the turbulent pressure in the IGM and hence study
its effect on gravitational clustering.

We will investigate these problems with cosmological hydrodynamic
simulation samples of the concordance model of $\Lambda$CDM. In \S
2, we present the equations governing the dynamics of vorticity and
rate of strain field. \S 3 gives a brief description of the
cosmological hydrodynamic simulation of the $\Lambda$CDM model. In
\S 4 we discuss the statistical properties of vorticity on large
scales. The nonthermal pressure of turbulent fluid and its effects
on clustering of the IGM are addressed in \S 5. We summarize the basic
results of the paper and give concluding remarks in \S 6.
Mathematical equations are given in the Appendix.

\section{Theoretical Background}

\subsection{Dynamical Equation of Vorticity}

The dynamics of a fluid is conventionally governed by a set of
equations for velocity and density fields $v_i(t,{\bf r})$,
$\rho(t,{\bf r})$ (Appendix \S A.1).  An alternative way is to
replace the velocity field by their spatial derivatives $\partial_i
v_j$. The velocity derivative tensor $\partial_iv_j$ can be
decomposed into a symmetric component
$S_{ij}=(1/2)(\partial_iv_j+\partial_jv_i)$ and an antisymmetric
component $(1/2)(\partial_iv_j-\partial_jv_i)$(Landau \& Lifshitz
1987). The former is the rate of strain and the latter is the
vorticity vector $\omega_i=\epsilon_{ijk}\partial_jv_k$, or
$\vec{\omega}=\nabla \times {\bf v}$, where $\epsilon_{ijk}$ is the
Levi-Civita antisymmetric symbol. For a cosmic baryon fluid
(IGM), the dynamical equation of vorticity $\vec{\omega}$ can be
derived from the Euler equation as (Appendix \S A.1 and \S A.2)
%eq1
\begin{equation}
\frac{D \vec {\omega}}{Dt}\equiv \partial_t {\vec \omega}
+\frac{1}{a}{\bf v}\cdot{\nabla}\vec{\omega}=\frac{1}{a}({\bf
S}\cdot {\vec\omega}-d {\vec\omega} +\frac{1}{\rho^2}\nabla \rho
\times\nabla p-\dot{a}\vec{\omega}),
\end{equation}
where $p$ is the pressure of the IGM, $a(t)$ is the cosmic factor,
$d=\partial_iv_i$ is the divergence of the velocity field, and the
vector $[{\bf S}\cdot {\vec\omega}]_i=S_{ij}\omega_j$. Defining a
scalar field as $\omega\equiv |{\vec \omega}|$, the dynamical
equation of $\omega$ is then
%eq2
\begin{equation}
\frac{D \omega}{Dt}\equiv \partial_t {\omega} +\frac{1}{a}{\bf
v}\cdot{\nabla}\omega=\frac{1}{a}\left [\alpha \omega-d\omega +
\frac{1}{\rho^2}\vec{\xi}\cdot (\nabla \rho \times\nabla p )
-\dot{a}\omega \right ],
\end{equation}
where $\vec{\xi}={\vec\omega}/\omega$, and
$\alpha={\vec\xi}\cdot({\vec\xi}\cdot\nabla){\bf v}$.

An essential feature of both eqs.(1) and (2) is that they are free
from the gravity of mass fields, therefore, the gravitational field
of both dark matter and the IGM cannot be a source of the vorticity.
Obviously, in the linear regime, only the last term of eqs.(1) and
(2) survives. This term is from the cosmic expansion, and makes the
vorticity decaying as $a^{-1}$. Thus, the vorticity of the IGM is
reasonably negligible in the linear regime.

Equations (1) and (2) show that if the initial vorticity is zero,
the vorticity will stay at zero in the nonlinear regime, provided that
the term $(1/\rho^2)\nabla \rho \times\nabla p$ is zero. This term,
called baroclinity, characterizes the degree to which the gradient
of pressure, $\nabla p$, is not parallel to the gradient of density,
$\nabla \rho$.

If the pressure of a baryon gas is a single-variable function of
density, e.g. there exists a determined relation for the equation of
state $p=p(\rho)$, the vector $\nabla p$ would be parallel to
$\nabla \rho$, and then $(1/\rho^2)\nabla \rho \times\nabla p=0$.
Therefore, vorticity cannot be generated even in the nonlinear regime
until the single-variable function or determined relation for $p=p(\rho)$ is
violated. Physically, once multi-streaming and turbulent flows have
developed, complex structures, like curved shocks, will lead
to a deviation of the direction of $\nabla p$ from that of $\nabla \rho$. In
this case, the $\rho-p$ relation cannot be simply given by an
single-variable function equation as $p=p(\rho)$ (He et al 2004) and the
baroclinity will no longer be zero.

The term  ${\bf S}\cdot{\vec\omega}$ on the right hand side of
eq.(1) accounts for stretch of vortices drived by strain. The
vorticity will be either amplified or attenuated by this term. Actually,
this point can be easily seen with eq.(2). If the coefficient
$\alpha$ is larger than zero, i.e. $\vec{\xi}$ is in the direction of
the eigenvector of tensor $\partial_jv_i$ with positive eigenvalue,  
the vorticity will grow at the rate of $\alpha \omega$. 
 Otherwise it would be attenuated. The term $-d\omega$ 
stands for expansion or contraction of vortices caused by the 
compressibility of baryon. Since divergence $d=\partial_jv_j$ is 
generally negative in regions of clustering, the term $-d\omega$ 
will lead to an amplification of vorticity in overdense regions.

\subsection{Vorticity Effect on IGM Clustering}

The effect of vorticity on the IGM clustering can be seen
from the dynamic equation of divergence d, which is an
indicator of clustering. The equation reads (Appendix \S A.3)
%eq3
\begin{eqnarray}
\frac{Dd}{Dt} & \equiv & \partial_td +\frac{1}{a}{\bf v}\cdot{\bf \nabla}d \\
\nonumber
 & = & \frac{1}{a}\left [\frac{1}{2}\omega^2-
S_{ij}S_{ij}-\frac{1}{\rho}\nabla^2 p
+\frac{1}{\rho^2}(\nabla\rho)\cdot(\nabla p) -\frac{4\pi
G}{a}(\rho_{tot}-\rho_0)-\dot{a}d  \right ],
\end{eqnarray}
where $\rho_{tot}$ is the total mass density including both CDM and baryon
and $\rho_{0}$ is its mean value.
A negative $d$ corresponds to a convergent flow (clustering), while
a positive $d$ means a divergent flow. As in the equation (2) for
vorticity, there is a term $-\dot{a} d$ coming from the cosmic
expansion that leads to dilution of $d$. However, different from
eqs.(1) and (2), the gravity effect $-4\pi G(\rho_{tot}-\rho_0)/a$,
$\rho_0$ acts as a source term in the divergence equation.  This
term leads to clustering in regions with
$\rho_{tot}>\rho_0$, and anti-clustering for
$\rho_{tot}<\rho_0$.

The term $(\nabla\rho)\cdot(\nabla p)/\rho^2$ will be nonzero even
when the IGM is barotropic, or the density-pressure relation is a
power law $p \propto \rho^{\gamma}$ and $\gamma>0$. The ratio of
this term to the gravity is roughly $\sim (t_{infall}/t_{sound})^2$,
where $t_{infall}\sim (G\rho)^{-1/2}$, and $t_{sound}\sim l/c_{s}$
with the typical scale of density variation $ l \sim
(\nabla\rho/\rho)^{-1}$ and the speed of sound $c_s\sim (\nabla
p/\nabla \rho)^{1/2}$ . The value of this ratio defines
roughly the Jeans criterion for gravitational instability. In
addition, the pressure term $-\nabla^2 p$ is compatible with
$(\nabla\rho)\cdot(\nabla p)/\rho^2$  and is likely to be positive
in overdense clustering regions. Hence, these two terms are from
thermal pressure to resist upon gravitational collapse.

Finally, we examine the effect of the first two terms, the strain
$S_{ij}S_{ij}$ and the vorticity $\frac{1}{2}\omega^2$, on
the right hand side of  eq.(3). For simplicity, we consider an
incompressible fluid in the absence of gravity. In this case, eq.(3)
simplifies to
%eq4
\begin{equation}
\nabla^2 p=-\rho\left (S_{ij}S_{ij}-\frac{1}{2}\omega^2 \right ).
\end{equation}
This is a typical Poisson equation for a scalar field of the pressure
$p$. Taking the similarity with the field equations in
electrostatics, the term on the right hand side of eq.(4),
$\mathfrak{Q}=\rho [S_{ij}S_{ij}-{1/2}\omega^2]$, mimics the
''charge" of a pressure field.  A positive  ''charge" produces an
attraction force that tends to drive overdense charge halos while a
negative "charge" yields a repulsive force that smear out the charge
accumulation. Back to the IGM flow, $\mathfrak{Q}$ plays the role of
 nonthermal pressure of turbulence (Chandrasekhar 1951, a, b;
Bonnazzola et al 1987). In regions with $\mathfrak{Q}<0$, the
turbulent pressure will prevent the IGM clustering. The sign of
$\mathfrak{Q}$ is actually determined by levels to which the
turbulence has developed(\S 5.2).

\section{Numerical Method}

To model the flow patterns of the IGM and dark matter fields, we use the
WIGEON code, which is a cosmological hydrodynamic/N-body code based
on the fifth-order WENO (weighted essentially non-oscillatory) scheme
(Feng et al. 2004). The WENO scheme uses the idea of
adaptive stencils in the reconstruction procedure based on the local
smoothness of the numerical solution to automatically achieve high order
accuracy and non-oscillatory property near discontinuities. Specifically,
WENO adopts a convex combination of all the candidate stencils, each being
assigned a nonlinear weight which depends on the local smoothness
of the numerical solution based on that stencil (Shu, 1998, 1999). For
more details, one can refer to Appendix A.4.

The WENO scheme has been successfully applied to hydrodynamic
problems containing turbulence (Zhang et al 2008), shocks and
complex structures, such as shock-vortex interaction (Zhang et al
2009), interacting blast waves (Liang \& Chen 1999; Balsara \& Shu
2000), Rayleigh-Taylor instability (Shi, Zhang \& Shu 2003). The
WENO scheme has also been used to simulate astrophysical flows,
including stellar atmospheres (del Zanna, Velli \& Londrillo 1998),
high Reynolds number compressible flows with supernova (Zhang et al.
2003), and high Mach number astrophysical jets (Carrillo et al.
2003). In the context of cosmological applications, the WENO scheme
has been proved to be especially adept at handling the Burgers' equation,
a simplification of Navier-Stokes equation,typically for modeling shocks
and turbulent flows (Shu 1999). This
code has also been successfully applied to reveal the turbulence
behavior of the IGM (He et al 2006, Liu \& Fang 2008, Lu et al
2009).

We evolve the simulation in the concordance model of a LCDM universe
specified by the cosmological parameters $(\Omega_{m},
\Omega_{\Lambda},h,\sigma_{8},\Omega_{b},n_{s},z_{re})=
(0.274,0.726,0.705,0.812,0.0456,0.96,11.0)$ (Komatsu et al., 2009).
The simulation is performed in a periodic cubic box of size of 25 $h^{-1}$
Mpc with a  $512^{3}$ grid and an equal number of dark matter particles, which
have mass resolutions $1.04 \times 10^{7} M_{\odot}$. To test the
convergence, we also run a low-resolution simulation with a $256^{3}$ grid
 and an equal number of dark matter particles in the same box. Radiative
cooling and heating are modeled using the primordial composition
$(X=0.76,y=0.24)$ and calculated as in Theuns et al.(1998). A
uniform UV background of ionizing photons is switched on at
$z_{re}$. Processes such as star formation and feedback due to
stars, galaxies and active galactic nuclei(AGN) are not included in
our simulation. The simulations start at redshift $z=99$, and
the snapshots are outputted at redshifts
$z=11.0,6.0,4.0,3.0,2.0,1.0,0.5,0.0$.

The tensor $\partial_iv_j$ of samples is then calculated by using a four-point
finite-difference approximation at the same grid that is used in the simulation.
For example, the partial derivatives of $\partial_y v_x$ at grid $l,m,n$ is
given by
\begin{equation}
\partial_yv_x(l,m,n)=\frac{2}{3}[v_x(l,m+1,n)-v_x(l,m-1,n)]
-\frac{1}{12}[v_x(l,m+2,n)-v_x(l,m-2,n)].
\end{equation}
Once all the partial derivatives of three velocity components are
generated, one can produce the fields of vorticity and the rate of
strain of these samples.

\section{Basic Properties of the IGM Vorticity}

\subsection{Configuration of the Vorticity Fields}

%fig1
\begin{figure}[htb]
\centering
\includegraphics[width=8.0cm]{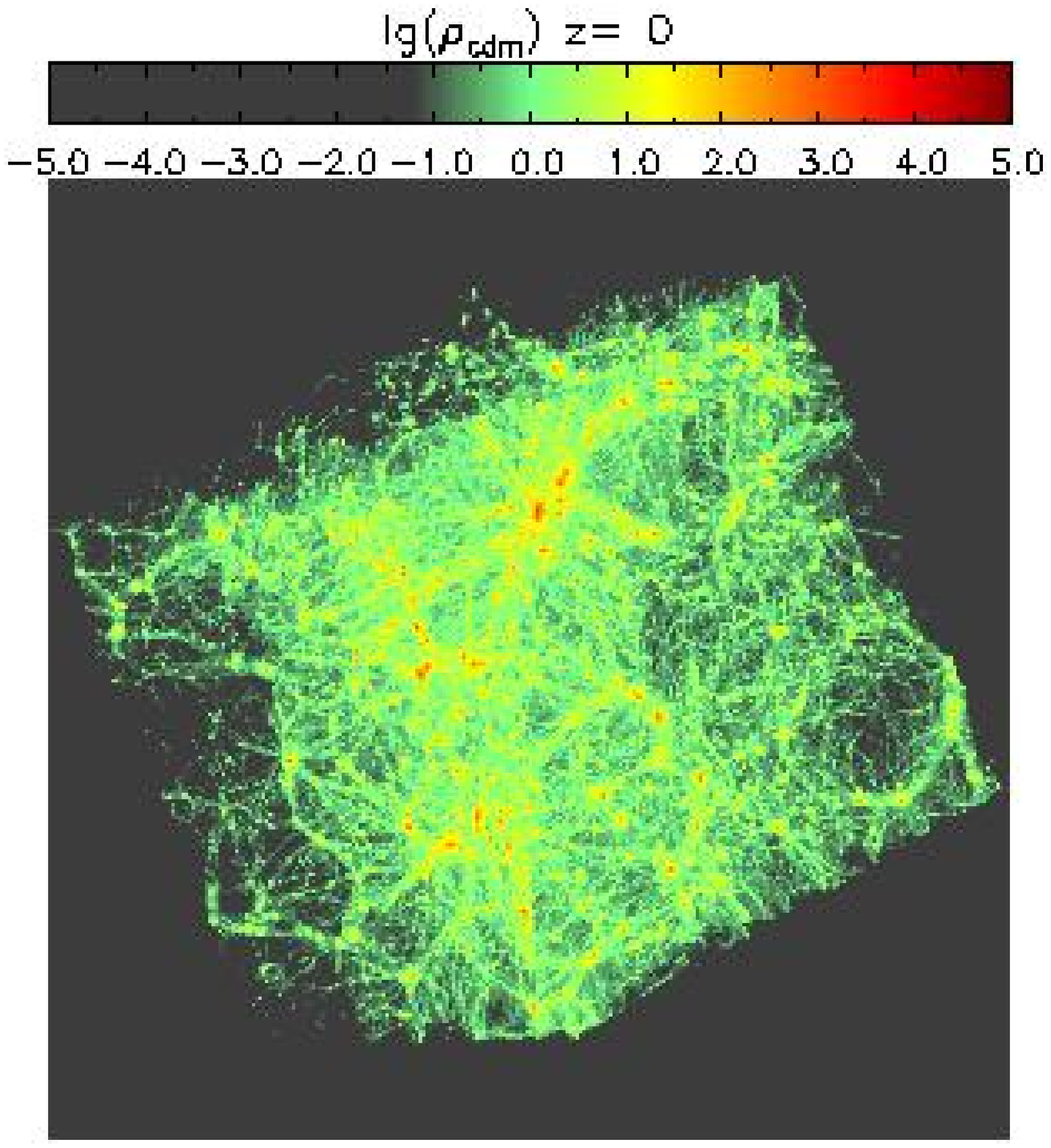}
\includegraphics[width=8.0cm]{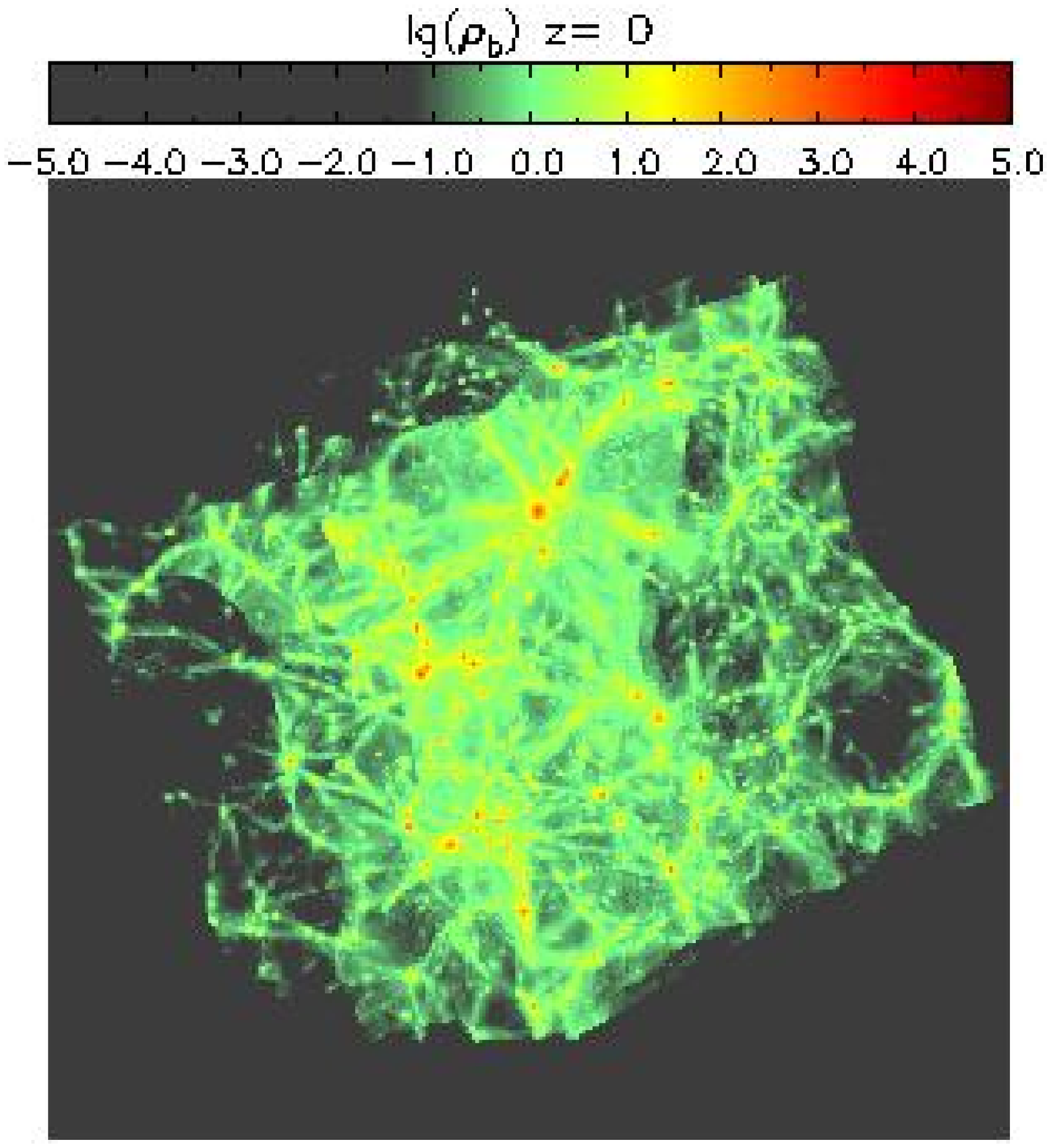}
\caption{3-D distribution of density fields of dark matter(left) and
baryon(right) in a periodic box size of $25 h^{-1}$ Mpc with a
$512^{3}$ grid.}
\end{figure}

Figure 1 visualizes 3-D density distributions of the dark matter
(left) and the baryon (right) respectively.  Figure 2 gives the 3-D distributions
of scalar field $\omega t$,$t$ is the cosmic time, at redshifts
$z=4$, 2 and 0. The dimensionless quantity $\omega t$ is to characterize the
typical number of eddy turnovers of vorticity
within the cosmic time. Figure 2 shows a strong evolution of the intensity
of vorticity with redshifts. The vorticity has not been well developed until
redshifts $z \sim 4$, but becomes significant at $z \sim 2$ which marks approximately
the onset of shocks and complex structures developed on the cosmic scales, and
matches with the typical formation history of dark matter halos of
galactic clusters (e.g. Bahcall \& Fan 1998).

%fig2
\begin{figure}[htb]
\centering
\includegraphics[width=5.0cm]{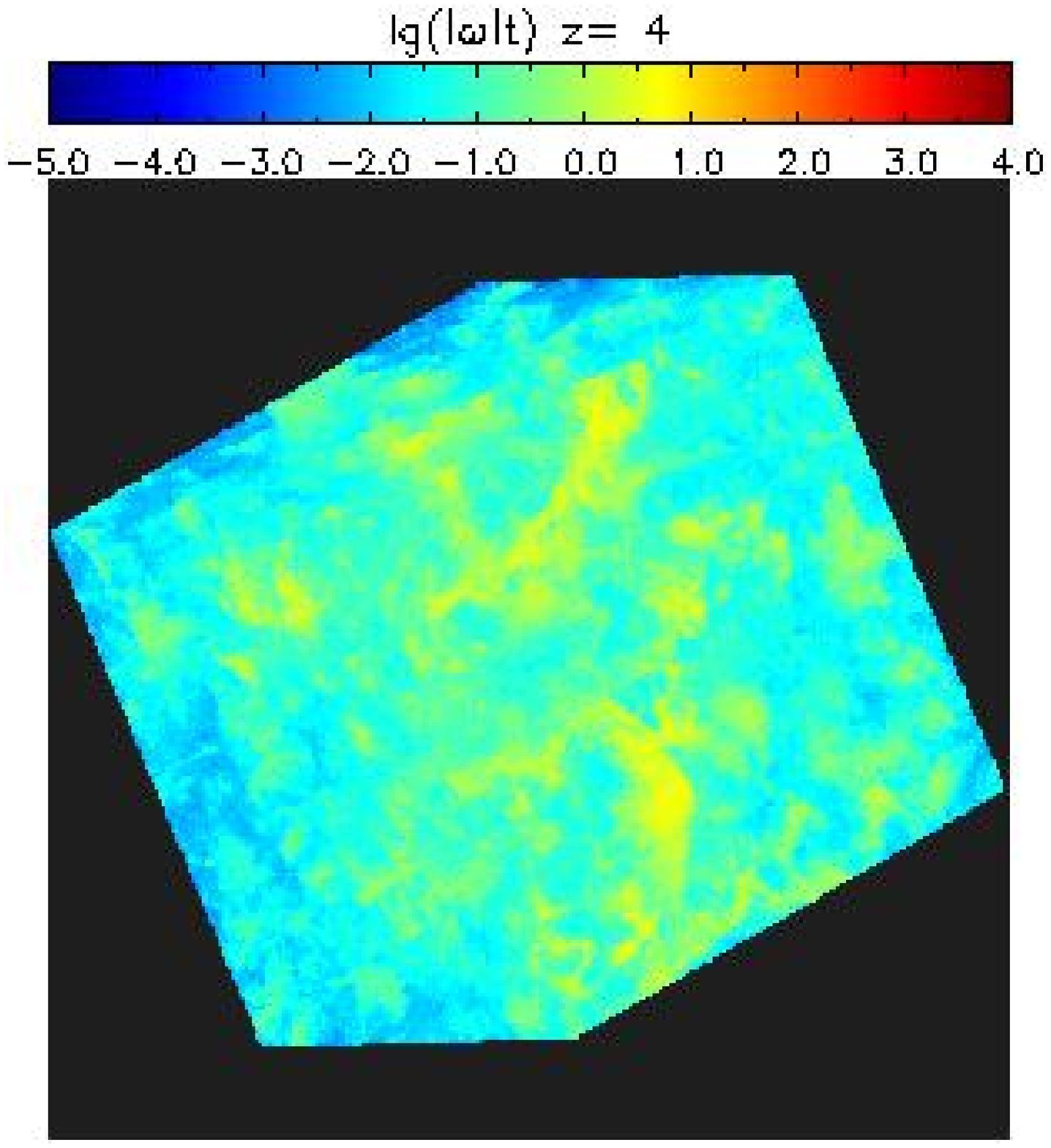}
\includegraphics[width=5.0cm]{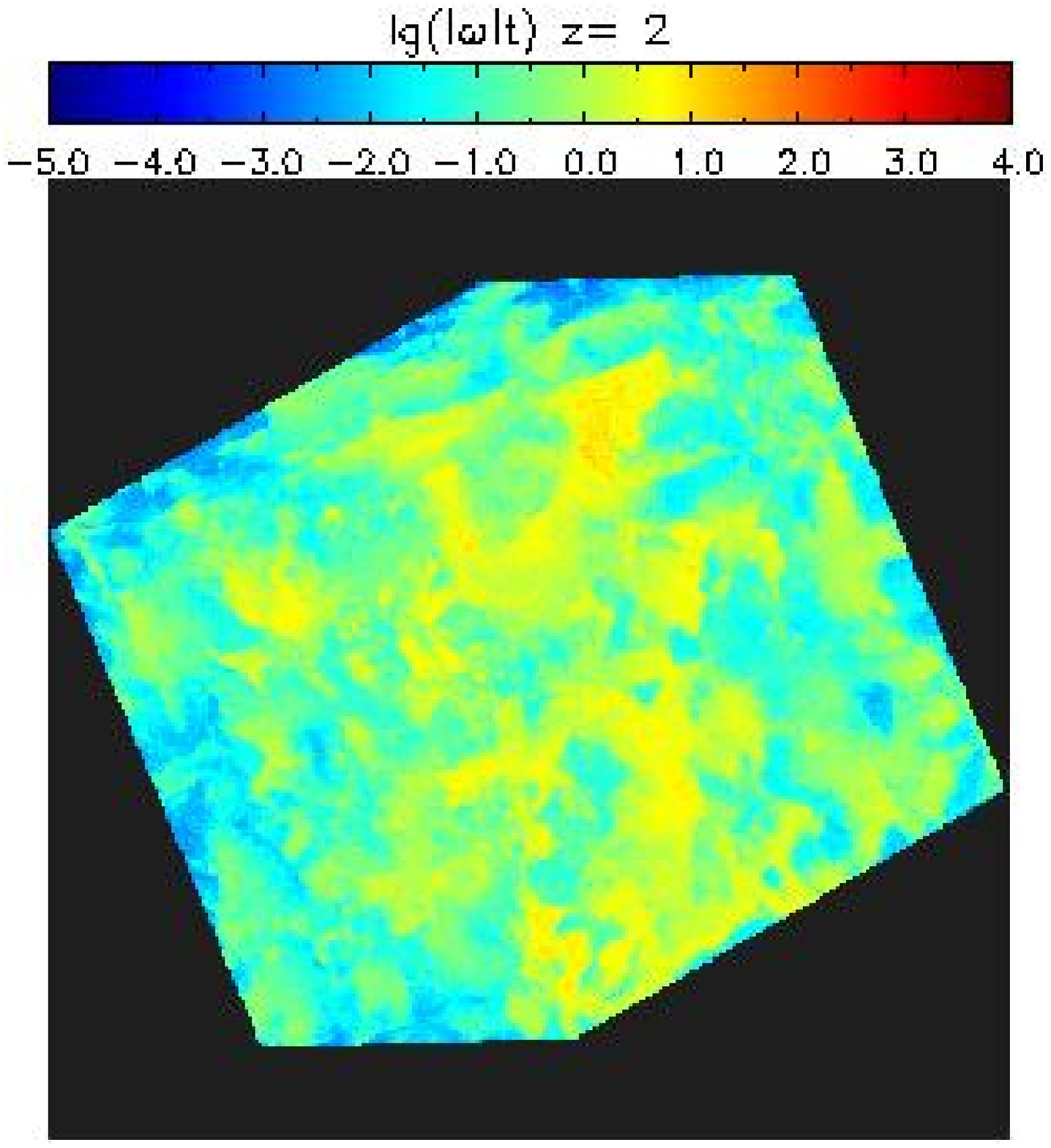}
\includegraphics[width=5.0cm]{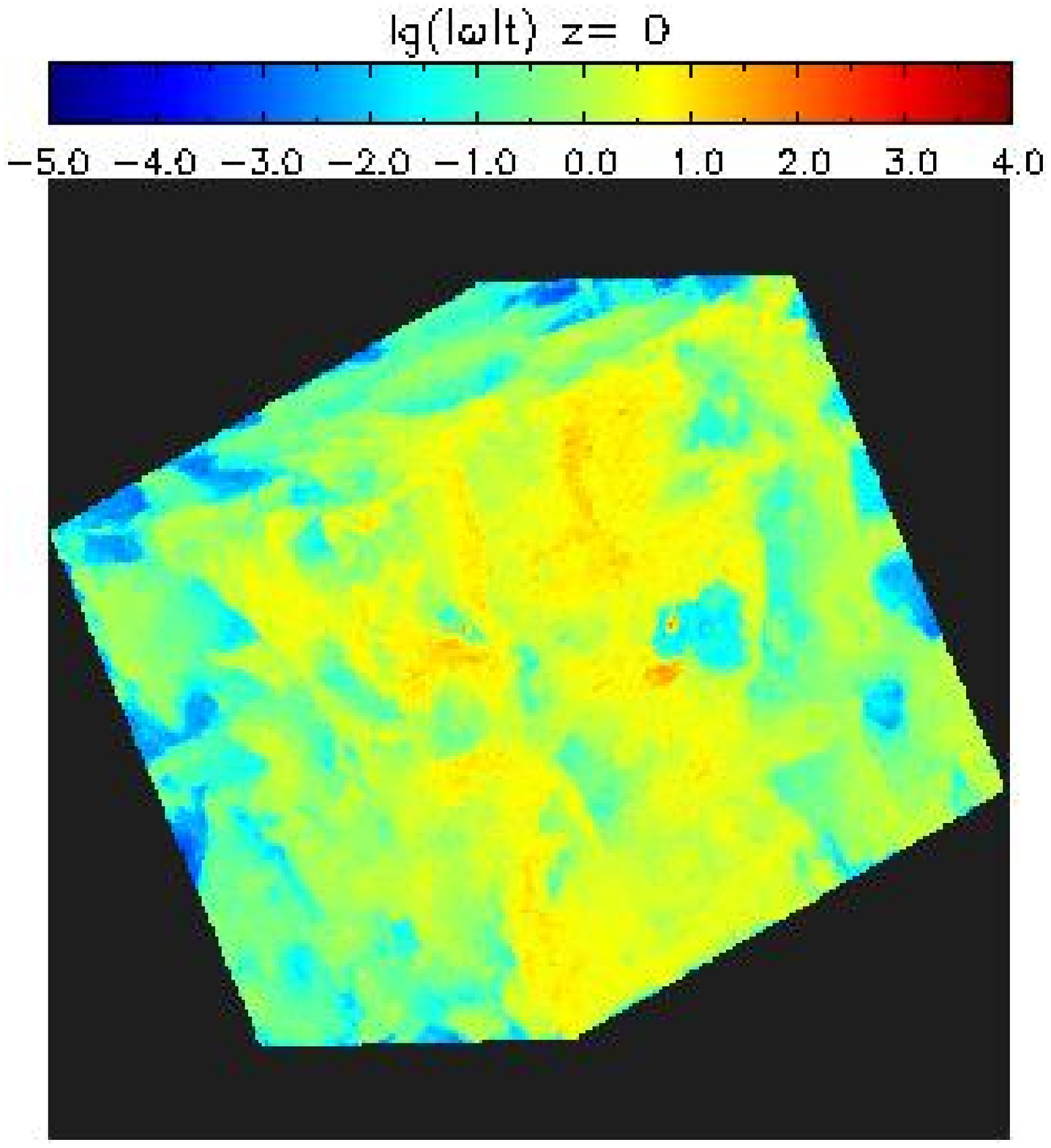}
\caption{3-D distribution of dimensionless scalar field $\omega t$ that
refers to the number of turnovers of vortices within the cosmic time, where $t$
 is the cosmic time, at redshifts $z=4$ (left), 2 (middle) and
0 (right).
 }
\end{figure}

The density fields (Figure 1) display the typical sheets-filaments-knot structures
on the cosmic scale. However, the spatial configuration of the vorticity field
looks quite different: it does not show any sheet-like or filamentary structure,
instead, looks like clouds with the comparable sizes of clusters. Although the
vorticity field is not associated with the fine structures of the underlying
density fields, the cloudy structures are most likely to occur around overdense
regions on large scales, and thus the vorticity field can be used to pick up
the coherent structures. These features can be more clearly illustrated by 2-D
distributions of $\omega t$ and vector projection of $\vec{\omega}$ shown in Figure 3,
where a slide of $25\times 25\times 0.1$ $h^{-3}$ Mpc$^3$ is taken. The distribution
of $\vec{\omega}$ shows a similar spatial pattern as the scalar quantity $\omega t$.

%fig3
\begin{figure}[htbp]
\begin{center}
\includegraphics[width=8.0cm]{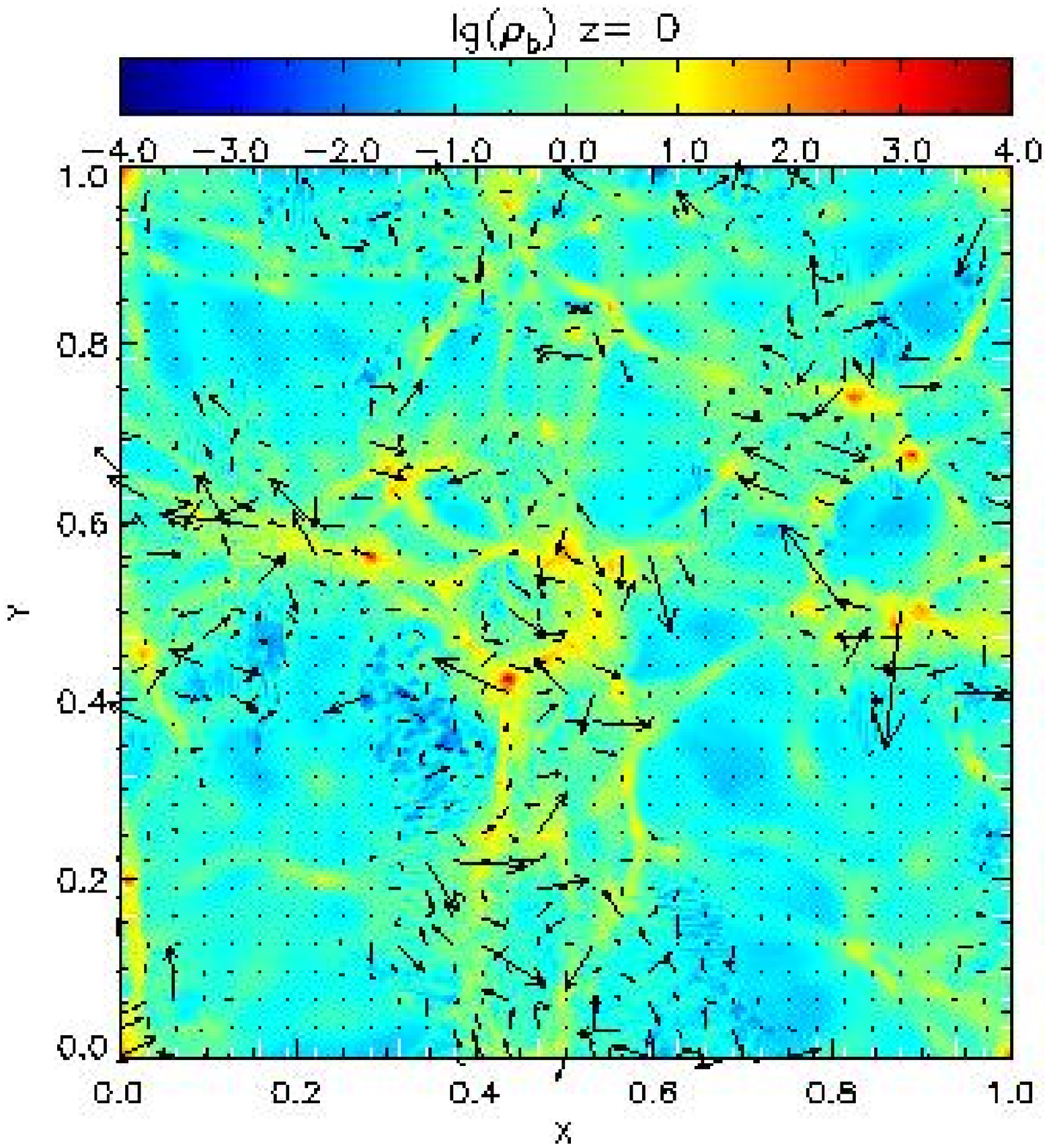}
\includegraphics[width=8.0cm]{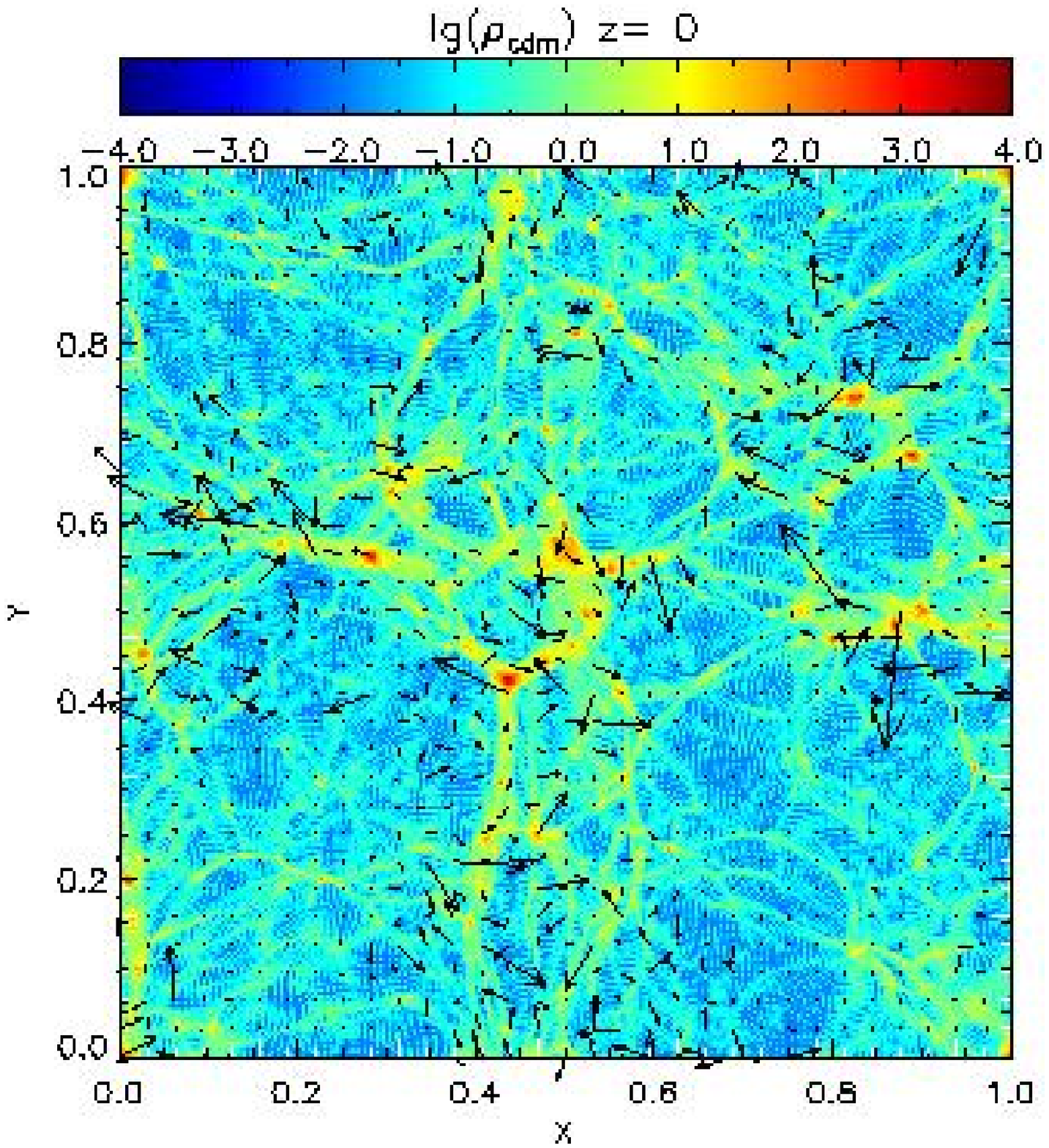}
\includegraphics[width=8.0cm]{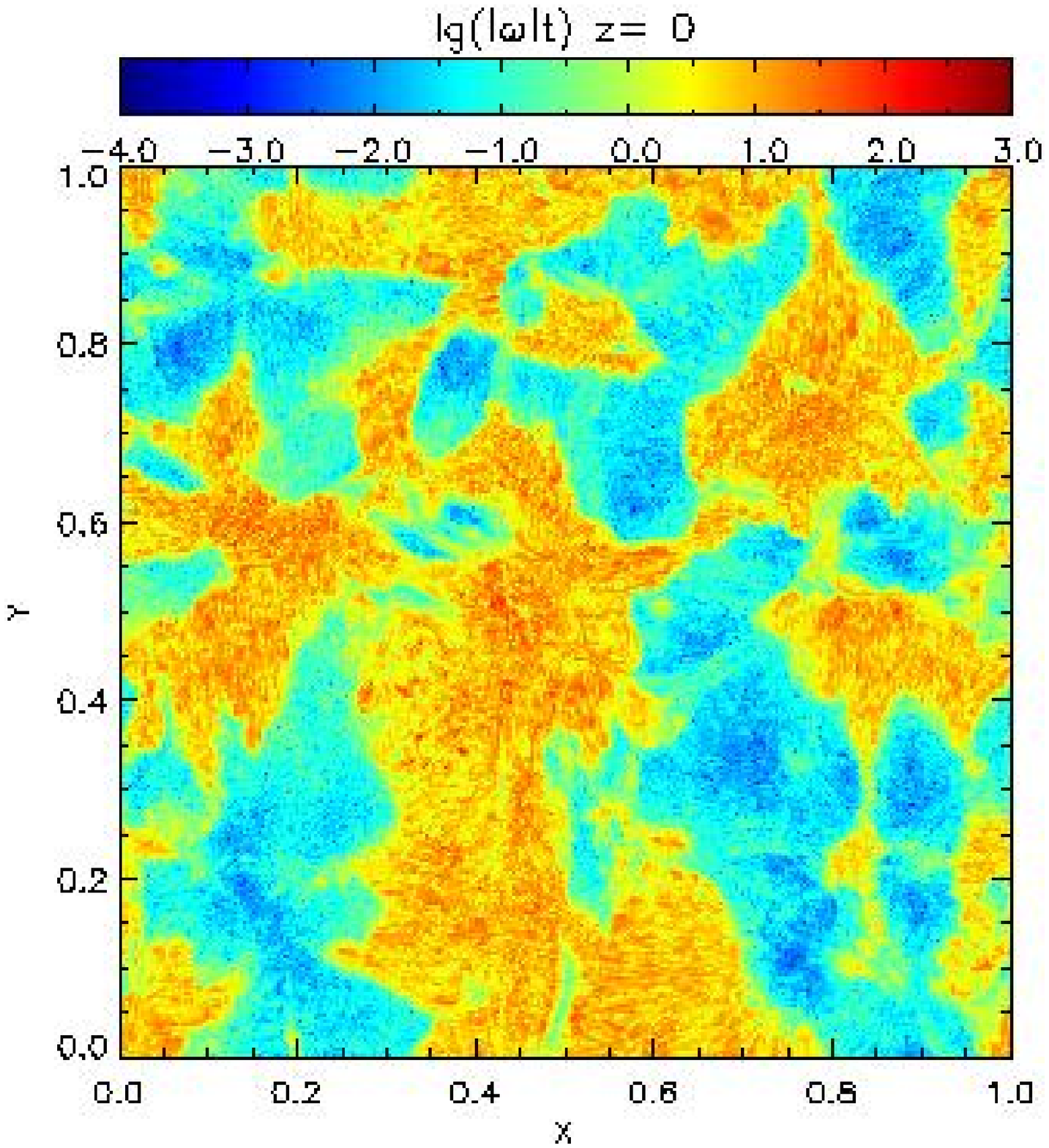}
\includegraphics[width=8.0cm]{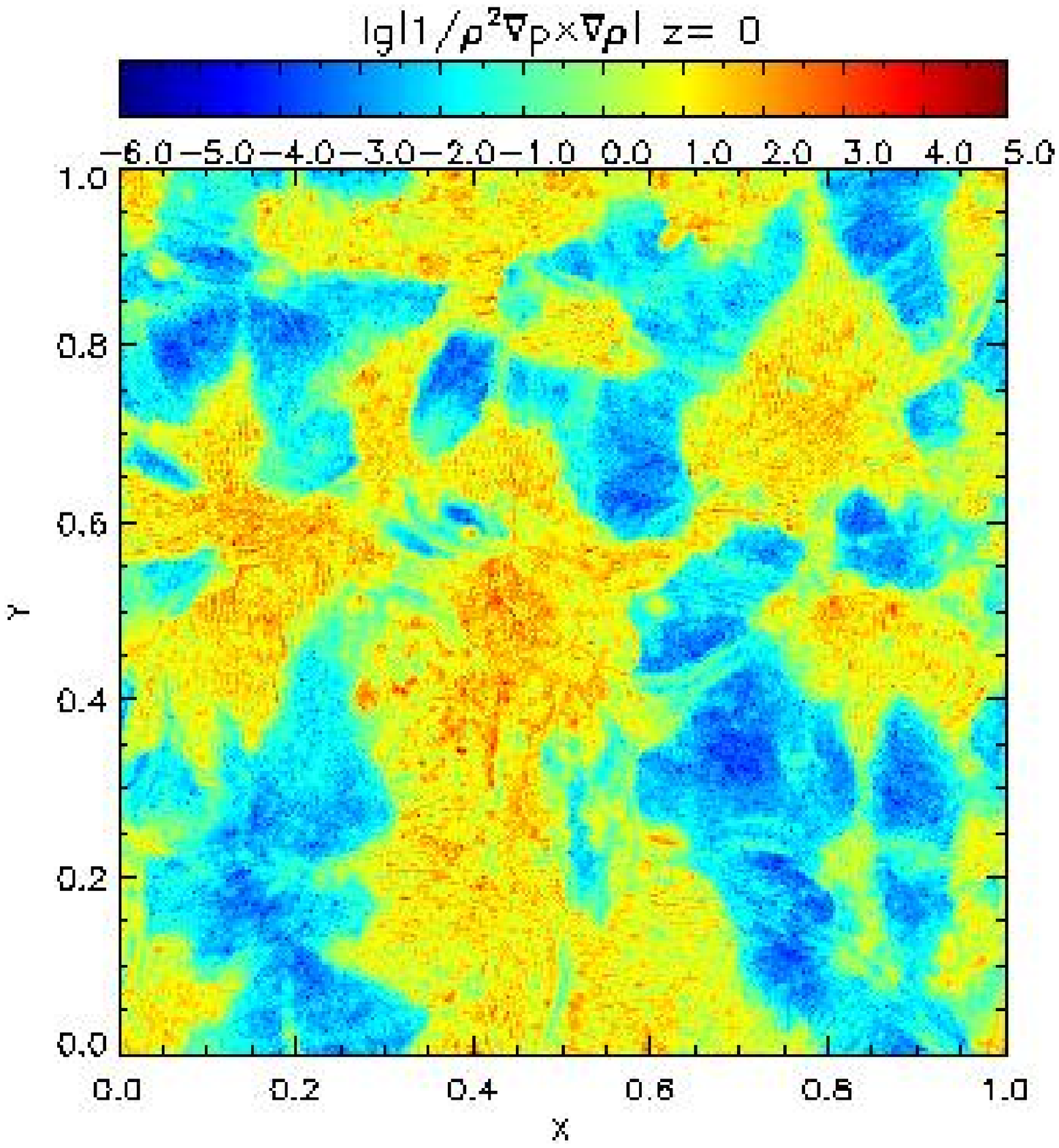}
\end{center}
\caption{Vorticity in a slide of $25\times 25\times 0.1$ $h^{-3}$Mpc$^3$. The
top two plots give vector fields of $\vec{\omega}$
against background of baryon density(top left) and dark matter density(top right).
The bottom left plot presents $\omega t$ in this slide
while the baroclinity field is given by the bottom right panel.}
\end{figure}

It is recalled that, in a  incompressible fluid, the evolution of vorticity is
driven dominantly by the amplification of the strain rate term ${\bf S}\cdot
{\vec\omega}$ [eq.(1)], which tends to stretch the vortical motion (Tanaka,
\& Kida, 1993; Constantin et al 1995) and produce filamentary(tubes) and/or
sheetlike structure in the vorticity field (e.g. She et al 1990).
As mentioned in \S 2.2, the strongest stretching is in the direction of $\vec\xi$,
which is parallel to the eigenvector of tensor $\partial_j v_i$ with
large positive eigenvalue. This mechanism distorts the vorticity field
and forms a tube-like network in the spatial configuration.

The IGM, however, is compressible as a fluid. The amplification due to the
strain rate ${\bf S}\cdot {\vec\omega}$ will be largely
canceled by the term $-d\omega$ [eq.(2)], which results in a
strong attenuation of vorticity in the direction parallel to the
eigenvector of the tensor $\partial_jv_i$ with  positive eigenvalues.
Consequently, the vector field $\vec{\omega}$ does not show any clear
sheetlike-filamentary structure.

Since vorticity is mostly attributed to the baroclinity $(1/\rho^2)\nabla
\rho \times\nabla p$, the distribution of vorticity should be
determined by the distribution of baroclinity. In figure 3, we also present the
baroclinity field in the same slide as that of $\omega t$. Clearly, both of
them show alike structures. It is noted that, similar to the vorticity, the
baroclinity can be strong even at low density regions, as shocks and complex
structures can be formed there (He et al 2004).

Nevertheless, there does not exist a linear mapping
between the $\omega t$ and the baroclinity. This is because the term
$|\alpha \omega -d\omega|$ is sometimes comparable with the baroclinity
$|(1/\rho^2)\vec{\xi}\cdot(\nabla\rho \times\nabla p)|$. Figure 4
gives a cell-by-cell comparison between $|\alpha \omega -d\omega|$
and $|(1/\rho^2)\vec{\xi}\cdot(\nabla\rho \times\nabla p)|$. In
average, the intensity of these two sources are almost of the same order. Thus,
the amplification and stretching by the rate-of-strain and divergence
cannot be ignored. It leads to the mapping between
the $\omega t$ and the baroclinity field deviating from a linear one.

%fig4
\begin{figure}[htb]
\begin{center}
\includegraphics[width=8cm]{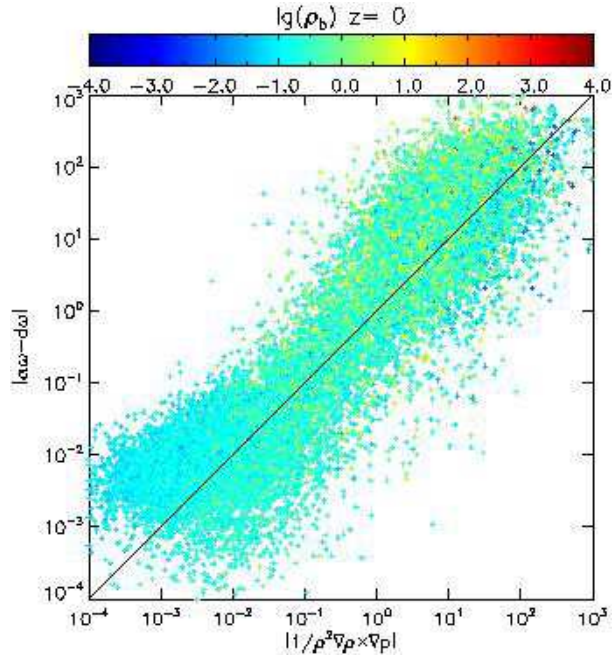}
\end{center}
\caption{A cell by cell comparison between the term $|\alpha \omega -d\omega|$, 
accounting for stretch in addition to expansion or contraction of vortices, and
baroclinity $|(1/\rho^2)\vec{\xi}\cdot(\nabla\rho \times\nabla p)|$, source of vorticity,
at redshift $z=0$.}
\end{figure}

\subsection{PDF of the Vorticity Fields}

%fig5
\begin{figure}[htb]
\begin{center}
\includegraphics[width=8cm]{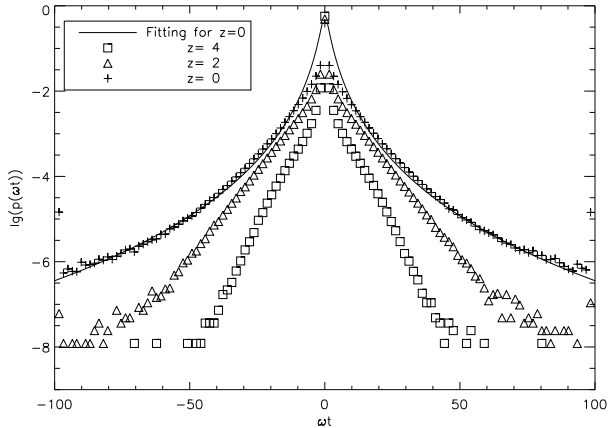}
\end{center}

\caption{The probability distribution function (PDF) of vorticity
$\vec{\omega_i}$ at redshifts $z=4$ (square), $z=2$ (triangle) and $z=0$
(cross).The solid line gives a log-normal fitting result for $z=0$.}
\end{figure}

We calculate the probability distribution function (PDF) of the three
components of the vector field $\vec{\omega_i}$ at redshifts $z=4$, 2, 0.
Giving that the vorticity field is isotropic, the PDFs of its three
components $\vec{\omega_i}$, $i=x,y,z$ should be statistically identical, which
is justified in our samples. We take an average over these three components
at these redshifts and give the results
in Figure 5. The PDF at present epoch exhibits a long tail, and can be
approximately fitted by a log-normal distribution as
%eq6
\begin{equation}
p(\omega t)d(\omega t) =  \frac{1}{\omega t\sqrt{2\pi \sigma^2}}
\exp\left [-\frac {1}{2}\left (\frac {\ln \omega t -\mu}{\sigma}\right)^2\right ]
d(\omega t)
\end{equation}
where the variance $\sigma = 0.98 ,\mu=0.37$,  which implies that the vorticity
field is intermittent, i.e. the probabilities of forming big vortical structures
are much larger than Gaussian fields.

It shows that the PDF of vorticity fields has been
always non-Gaussian since redshift $z \sim 4$, which is remarkably
different from the velocity field of the IGM. The PDF of the velocity
and pairwise velocity fields of dark matter and the IGM are Gaussian at
high redshifts, corresponding to the linear phase of evolution (Yang, et al 2001).
The evolution of the IGM vorticity field does not undergo a
 linear and Gaussian phase over cosmic times, since the vorticity can only be
produced via  nonlinear evolution.  In this sense, the vorticity field is more
effective than the velocity field to track the nonlinear evolution of the IGM.

Another interesting feature indicated in Figure 5 is that
the PDF at high redshifts is approximately of exponential,  and evolves into
log-normal distribution at later phase.
Thus,  the PDF at different redshifts cannot be converted to each other by
a scaling transformation.
It implies that the turbulence experiences a strong nonlinear evolution,
which will be revisited in next subsection.

\subsection{Power Spectra of Velocity and Vorticity Fields}

In a statistically homogeneous fluid, one can define the spectrum tensors
$\Phi_{ij}({\bf k})$ and
$\Omega_{ij}({\bf k})$ as the Fourier counterparts of the
two-point correlation tensors of velocity
$\langle v_i({\bf x+r})v_j({\bf x})\rangle$ and vorticity
$\langle \omega_i({\bf x+r})\omega_j({\bf x})\rangle$,

%eq7,
\begin{equation}
\Phi_{ij}({\bf k})=\frac {1}{(2\pi)^3}\int
\langle v_i({\bf x+r})v_j({\bf x})\rangle e^{-i{\bf k}\cdot {\bf r}}d{\bf r}
\end{equation}
%eq8
\begin{equation}
\Omega_{ij}({\bf k})=\frac {1}{(2\pi)^3}\int
\langle \omega_i({\bf x+r})\omega_j({\bf x})\rangle e^{-i{\bf k}\cdot {\bf r}}d{\bf r}.
\end{equation}
respectively, where $\langle ...\rangle$ denotes  average over spatial
coordinates ${\bf x}$.

For a homogeneous turbulence, we have (Batchelor, 1959)
%eq9
\begin{equation}
\Omega_{ij}({\bf k})=[\delta_{ij}k^2-k_ik_j]\Phi_{ll}
({\bf k})-k^2\Phi_{ij}({\bf k}),
\end{equation}
and hence,
%eq10
\begin{equation}
\Omega_{ii}({\bf k})=k^2\Phi_{ii}({\bf k}).
\end{equation}
The power spectra of velocity and vorticity fields are defined respectively as
%eq11
\begin{equation}
P_v(k)=\int \frac{1}{2}\Phi_{ii}({\bf k})\delta(|{\bf k}|-k)d{\bf
k}; \hspace{3mm}
P_{\omega}(k)=\int \frac{1}{2}\Omega_{ii}({\bf
k})\delta(|{\bf k}|-k)d{\bf k}.
\end{equation}
Combining eqs. (10) and (11) yields
%eq12
\begin{equation}
P_{\omega}(k)=k^2P_v(k).
\end{equation}
This relation is an important property of homogeneous turbulence
(Batchelor, 1959),  and can be used to measure the developed level
of turbulence. If the velocity and vorticity fields of a fluid
satisfy the relation given by eq.(12), it should be in the state of
fully developed  homogeneous turbulence. Otherwise, it would be less
developed.

%fig6
\begin{figure}[htbp]
\begin{center}
\includegraphics[width=8cm]{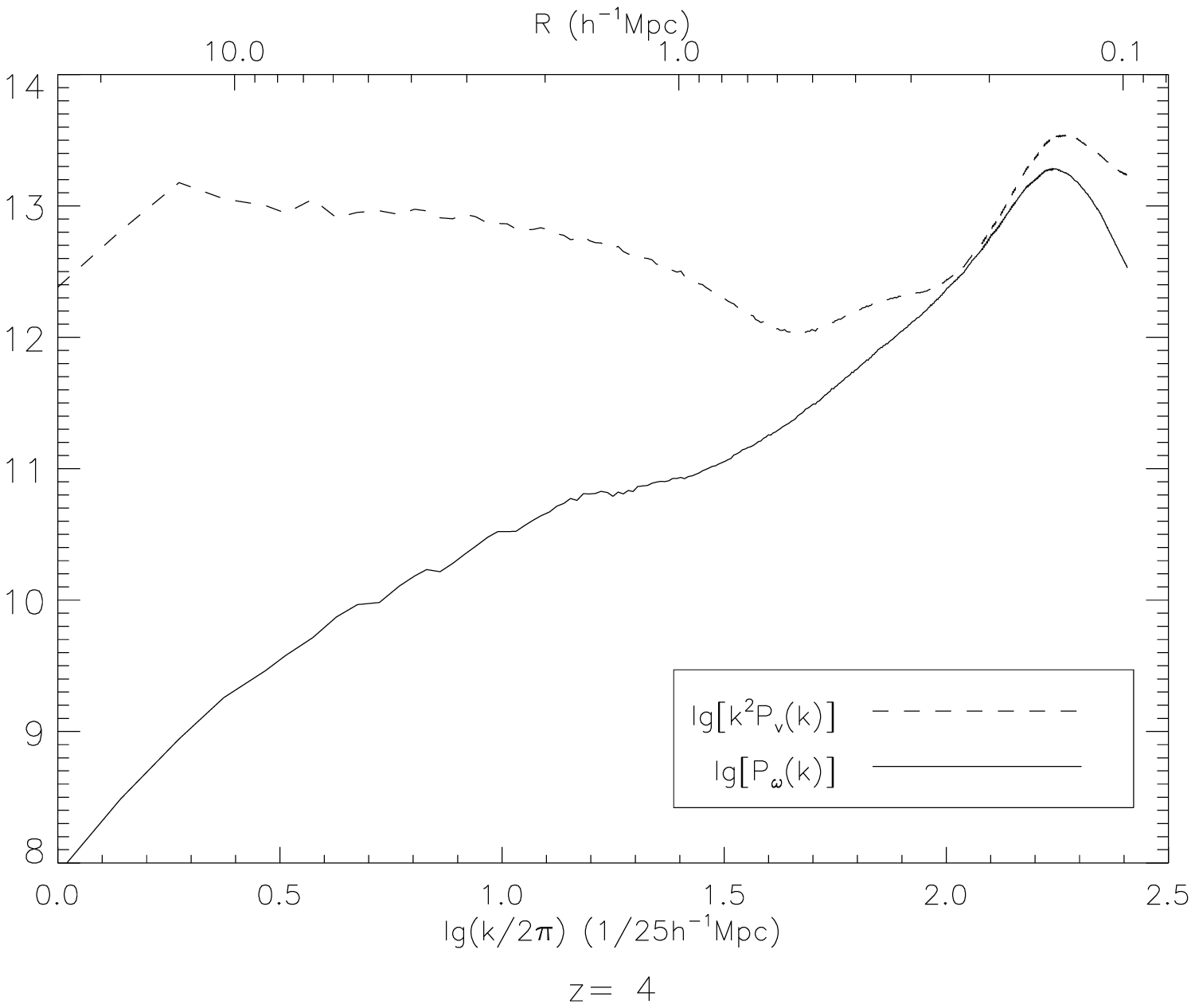}
\includegraphics[width=8cm]{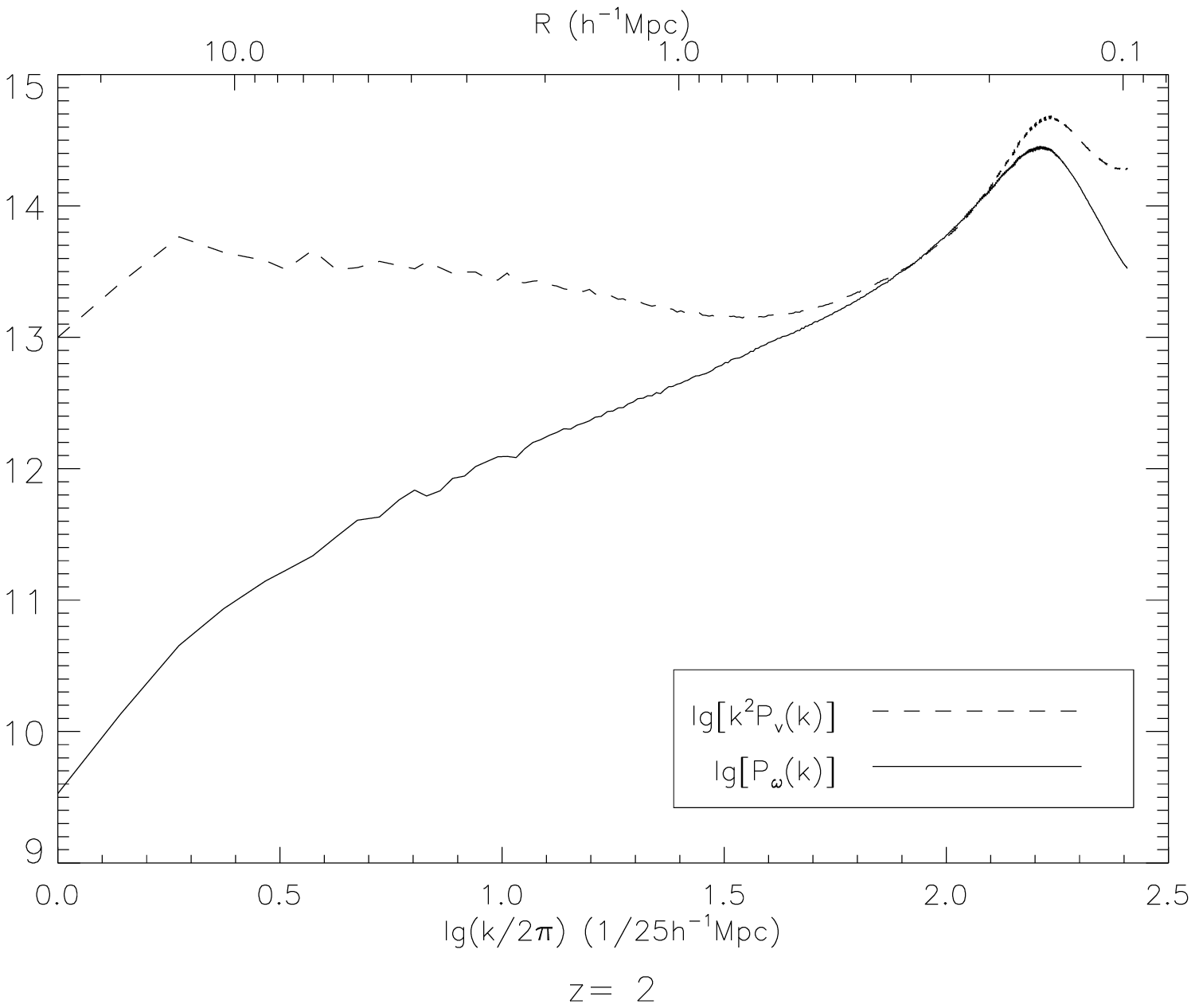}
\includegraphics[width=8cm]{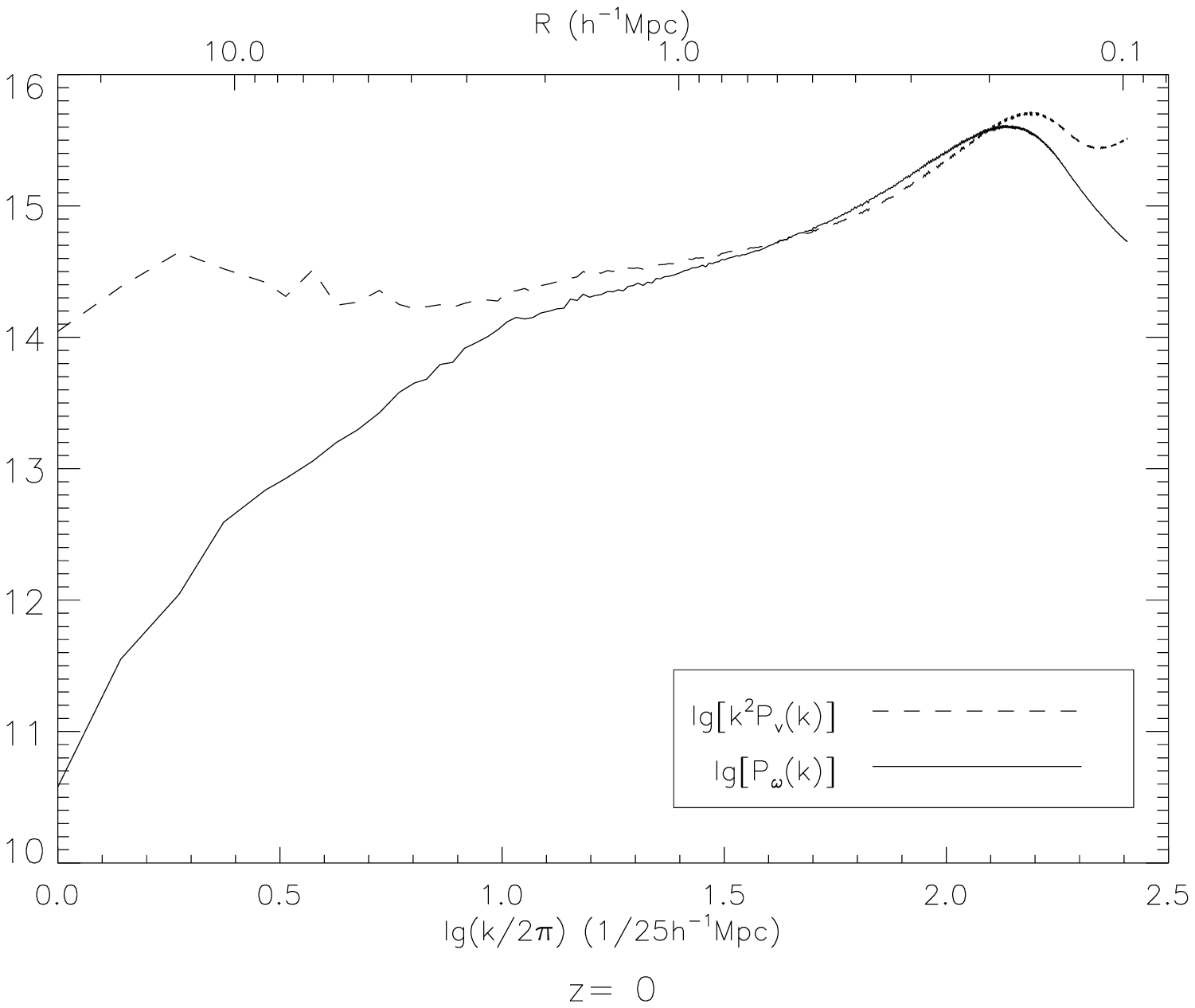}
\includegraphics[width=8cm]{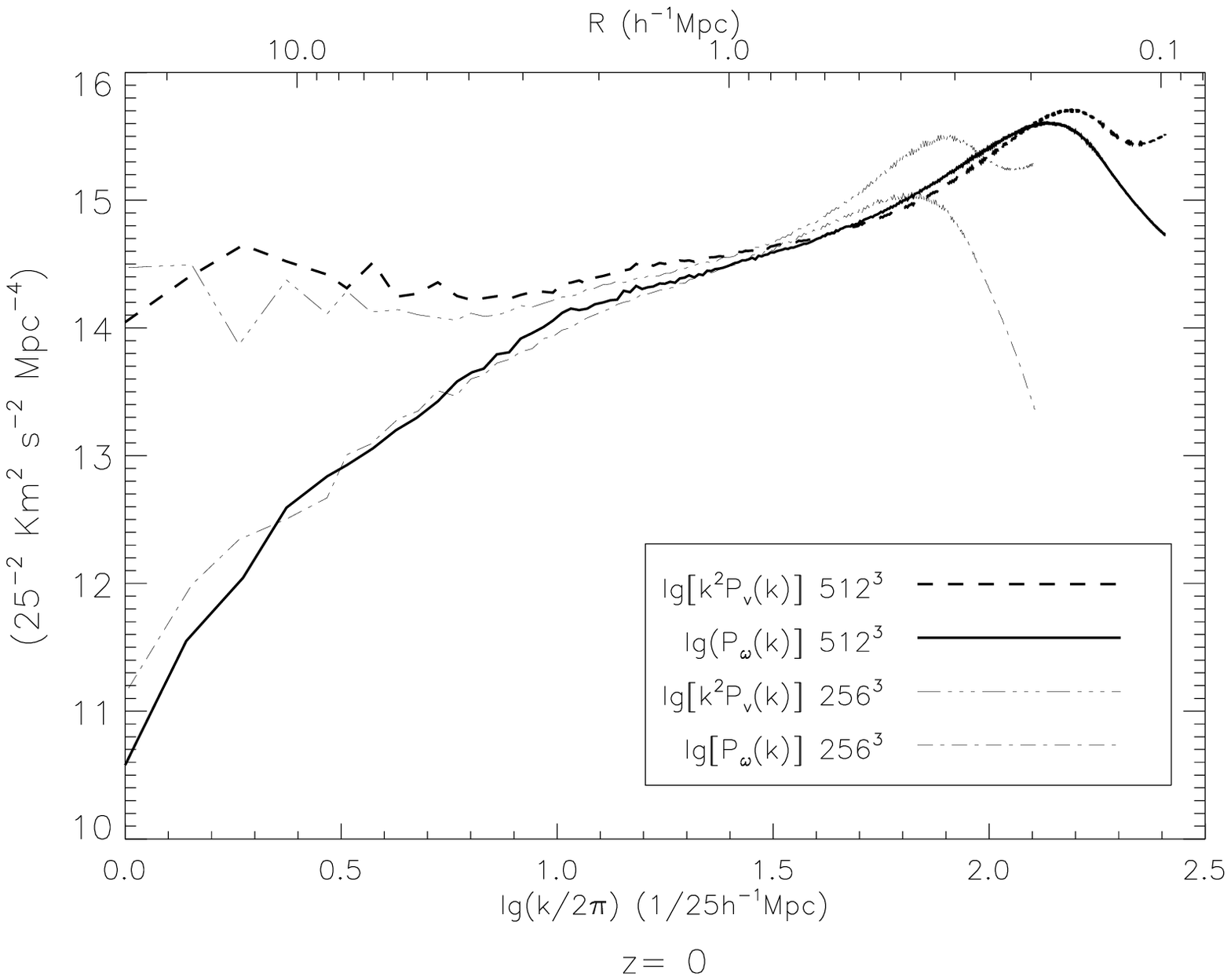}
\end{center}
\caption{The power spectra $P_{\omega}(k)$ and $k^2P_{v}(k)$ at redshifts $z=4$
(top left), 2 (top right ) and 0 (bottom left) from $512^{3}$ simulation. The bottom
right plot gives a resolution comparison of these two terms at redshift $z=0$.}
\end{figure}

Figure 6 compares the power spectra $P_{\omega}(k)$ with $k^2P_{v}(k)$ at $z=4$
(top left), 2 (top right) and 0 (bottom left), respectively. It shows that at
high redshift $z=4$, the power spectrum $P_{\omega}(k)$ is much less than
$k^2P_{v}(k)$ at almost all scales, which means that not all, actually only a
small part, of the fluctuations of velocity field can be related to the random
field of vorticity, and the turbulence is less developed by that time.
While evolving to redshift $z\sim 2$, the turbulence is developed starting
from the small scale 0.2$h^{-1}$ Mpc and up to 0.8$h^{-1}$ Mpc.  At the present
time, $z=0$,  the turbulence is fully developed and extended to the scale
3$h^{-1}$Mpc, the typical scale of a cluster. The deviations of
$P_{\omega}(k)$ from  $k^2P_{v}(k)$ on scales less than 0.2 $h^{-1}$ Mpc are
probably due to the energy dissipation of turbulence to thermal energy, or the
virialization, on small scales. A panel of these two terms in the simulation run
of lower resolution, $256^{3}$, is also presented in Fig. 6 and provides a convergence test
of the resolution effect. It shows that the resolution does affect the lower end
of turbulent scale as a result of dissipation. However, the turbulence on large
scale is resolution converged.

Figure 6 also shows that the variance of velocity field on large
scales is remarkably larger than that of vorticity field, especially
at high redshifts. It indicates that the variance on large scales is
not from the turbulent motion of the IGM and probably from bulk
motion, which is due mainly to the falling into gravitational well.
Therefore, to identify the variance of a velocity field as the signature of
turbulence (e.g. Iapichino \& Niemeyer 2008) may be questionable
even on scales of clusters, as they generally contain many
substructures at redshifts less than 2.

%fig7
\begin{figure}[htb]
\begin{center}
\includegraphics[width=6.0cm]{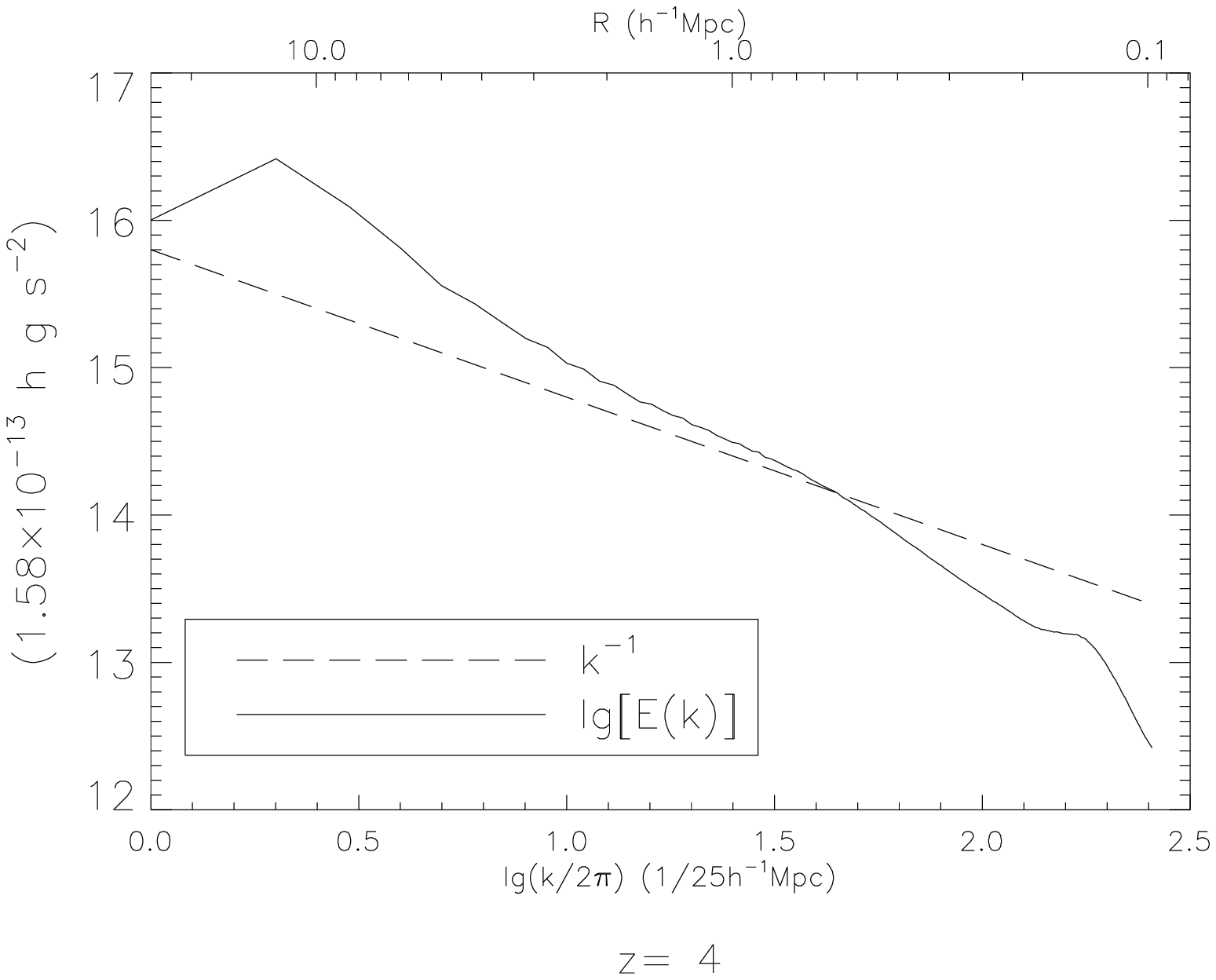}
\includegraphics[width=6.0cm]{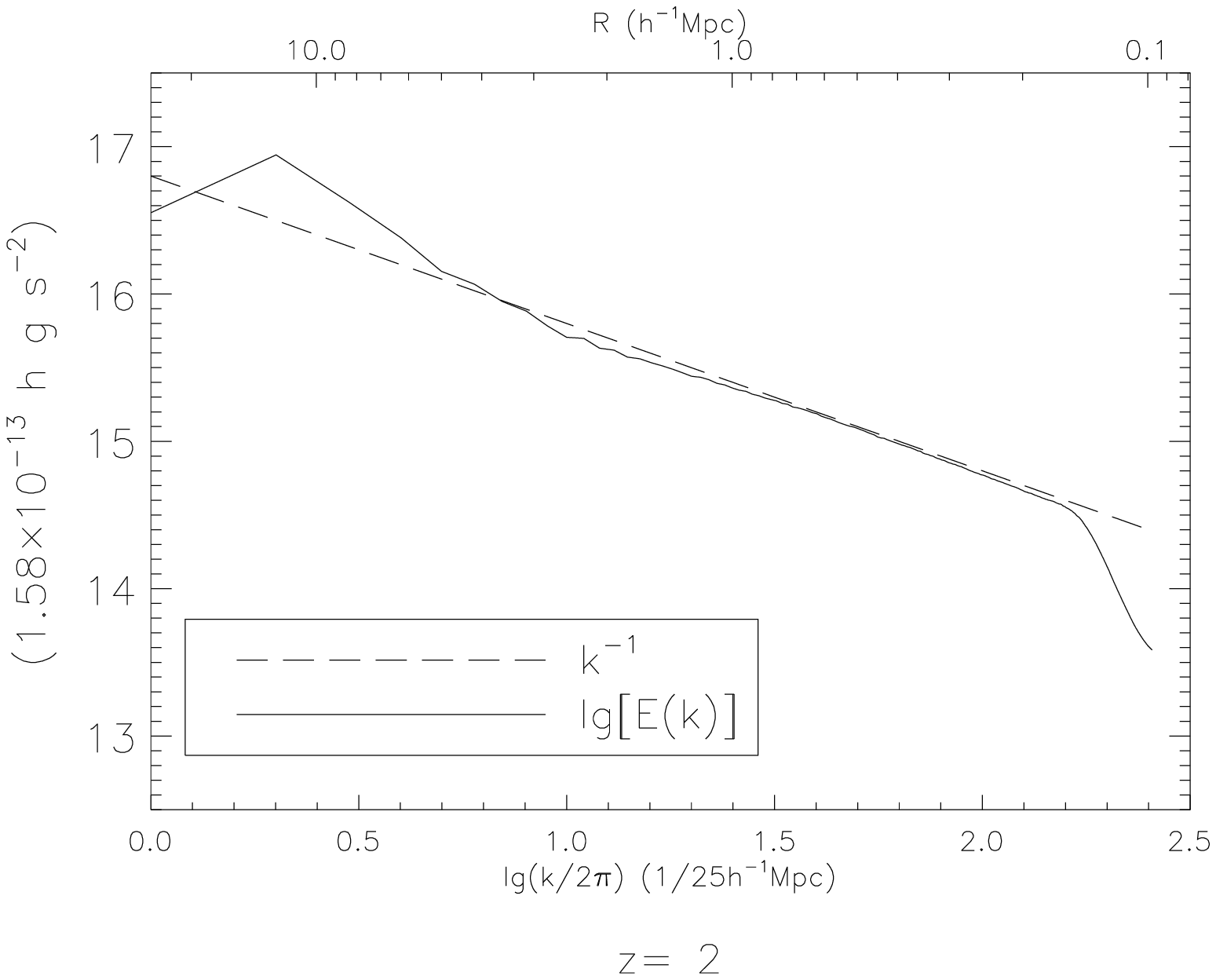}
\includegraphics[width=6.0cm]{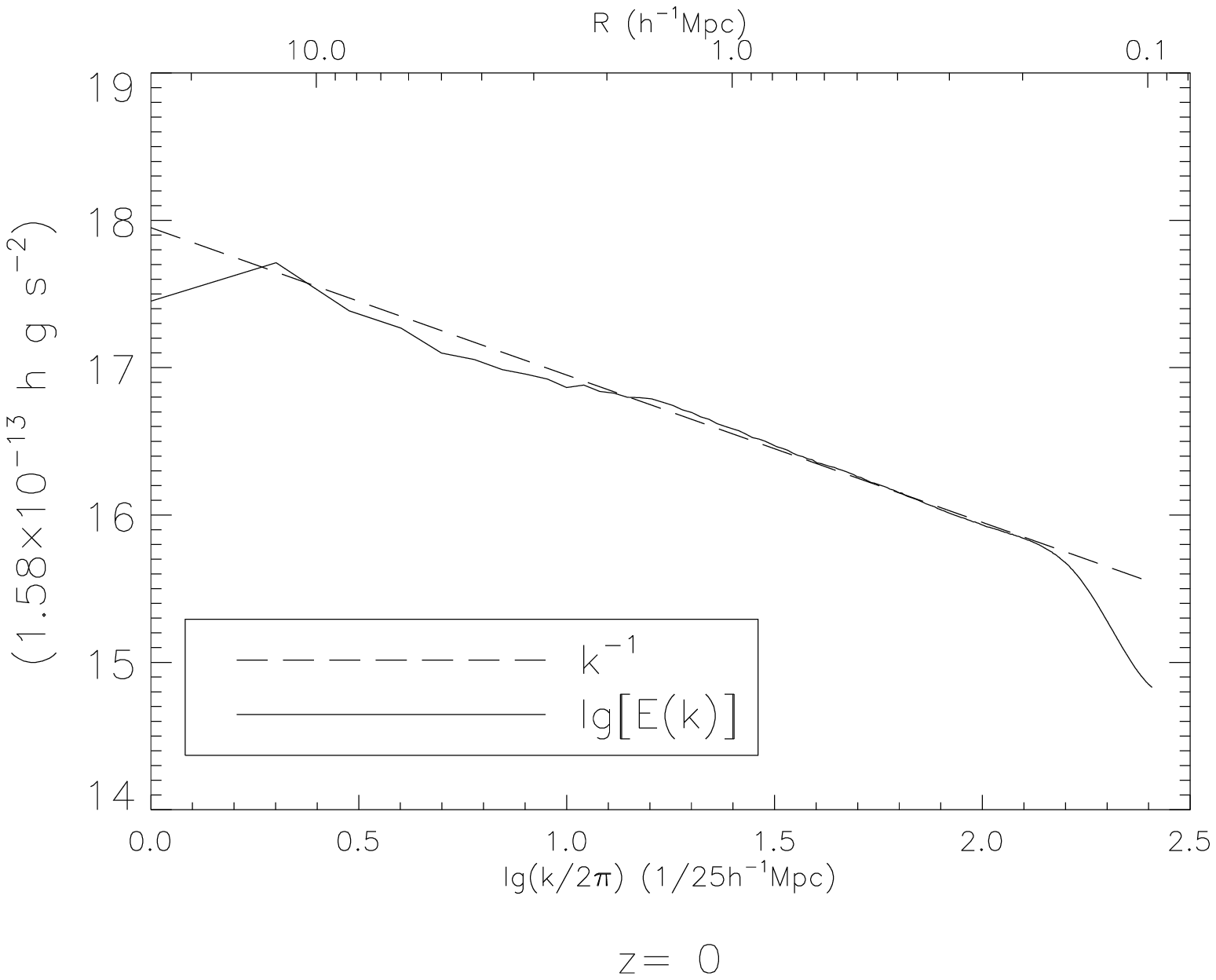}
\end{center}
\caption{The power spectrum of kinetic energy(solid line) at redshifts $z=4$
(left), $2$ (middle) and $0$ (right). A power law $k^{-1}$ (dashed line) is
used to fit the power spectrum.}
\end{figure}

We can also explore the evolution of turbulence with the spectrum of mean kinetic
energy density $E(k)$ defined by
$\int_0^\infty E(k)dk= 1/2\langle\rho({\bf r}) v^2({\bf r})\rangle$.
The energy spectra $E(k)$ at redshifts $z=4$, 2 and 0 are shown in
Figure 7. The energy spectra can be approximately fitted by a power law
$k^{-\alpha}$ with  $\alpha =1$ in the scale range of 0.15 - 3 $h^{-1}$ Mpc for
$z=2$ and 0.15 - 10 $h^{-1}$ Mpc for $z=0$. These scale ranges are
larger than that given by the power spectrum of velocity and vorticity. This is
probably because the turbulent flow is strong in high density areas.
Figure 7 shows that the energy spectrum becomes very steep at scales less than
0.15 $h^{-1}$ Mpc because of the dissipation on small scales. The energy spectrum
at $z=4$ cannot be fitted with the power law of $k^{-1}$. It indicates that
turbulence has not yet developed by then. Turbulence is effective at transferring
kinetic energy on large scales to small one. Therefore, it leads to the power 
spectrum at $z<4$ to be more flat than that of $z=4$.

\section{Effects of Turbulent IGM on Structure Formation}

\subsection{Non-thermal Pressure}

An early attempt of including the effect of turbulent motions into gravitational
collapsing processes was made by Chandrasekhar (1951). In his quantitative theory,
he investigated the effect of micro turbulence in the subsonic regime. If turbulence
is statistically homogeneous,  it will contribute an extra pressure
$p_{tub}=\rho\langle v^2\rangle$ on large scales. In the linear regime, Chandrasekhar
derived a dispersion relation by introducing an effective sound speed
$c^2_{s,eff}=c_s^2+ (1/3) \langle v^2\rangle$  where $\langle v^2\rangle$ is
the rms velocity dispersion due to turbulent motion. Obviously, the turbulence
will slow down, or even halt the gravitational collapsing.

Chandrasekhar's result had been improved by a more elaborate
investigation (Bonazzola et al. 1992) , in which the scale
dependence of the turbulent energy was taken account in the analysis
of system instability. Actually, the gravitational instability on a
scale $R$ will not be affected by fluctuation modes with
wavelengths larger than $R$, and the fluctuation of velocity on the
scales $k< 2\pi /R$ do not contribute to the turbulent pressure for
resisting on gravitational collapsing on scales that larger than
$R$. Quantitatively, the turbulent pressure on the scale $R$ can be
estimated by (Bonazzola et al 1987)
%eq8
\begin{equation}
p_{tur}(k_{R})=\int_{k_{R}}^{k_{max}} E(k) dk,
\end{equation}
where $E(k)$ is the turbulent power spectrum,  $k_{R}=2\pi/R$, and
${k_{max}=2\pi/l_{diss}}$ is the wavenumber
corresponding to the minimal scale $l_{diss}$ below which the turbulence
decays due to energy dissipation or virialization.

According to the results presented in \S 4.3, the turbulence is fully
developed on scales from 0.2
$h^{-1}$ Mpc up to 3 $h^{-1}$ Mpc since redshift $z\sim 2$. The
direct outcome of the turbulence on those scales is expected to
alter significantly the hydrostatic equilibrium state of the IGM or
the process of structure formation.  The turbulent pressure
$p_{tur}(k_{R})$ as a function of $k_{R}$ is shown in Figure 8,
where $l_{diss} =0.2$h$^{-1}$Mpc inferred from the power spectrum
analysis in \S 4. 3.  In practical calculation, since the energy
spectrum $E(k)$ declines fast beyond 0.2 $h^{-1}$ Mpc (Fig. 7), one
can take $k_{max}$ going to infinity safely. Since we have approximately
$E(k)\propto k^{-1}$, $p_{tur}(k_{R})$ given by eq.(13) is weakly
dependent on $k$. We also show the energy spectra $E(k)$ in Figure 8.

Using the power spectra measured in Figure 7,  the turbulent pressure is
estimated to be $1.5\times 10^{-17}$ g cm$^{-1}$s$^{-2}$ at $z=0$. According to
$p/\rho=RT/\mu$, the effective temperature due to turbulent pressure
is about $1.0\times 10^6$ K in regions of mean overdensity and 
$1.0\times 10^5$ K in regions of 10 time mean overdensity. Deduced from
Ly$\alpha$ forests of quasars, the temperature of IGM at
$\rho_b\simeq 1-10 \rho_{b,0}$ is about $2\times 10^4$ K. Therefore,
the nonthermal pressure of the turbulent flow could be larger
than the thermal pressure of the IGM.

%fig8
\begin{figure}[htb]
\begin{center}
\includegraphics[width=8cm]{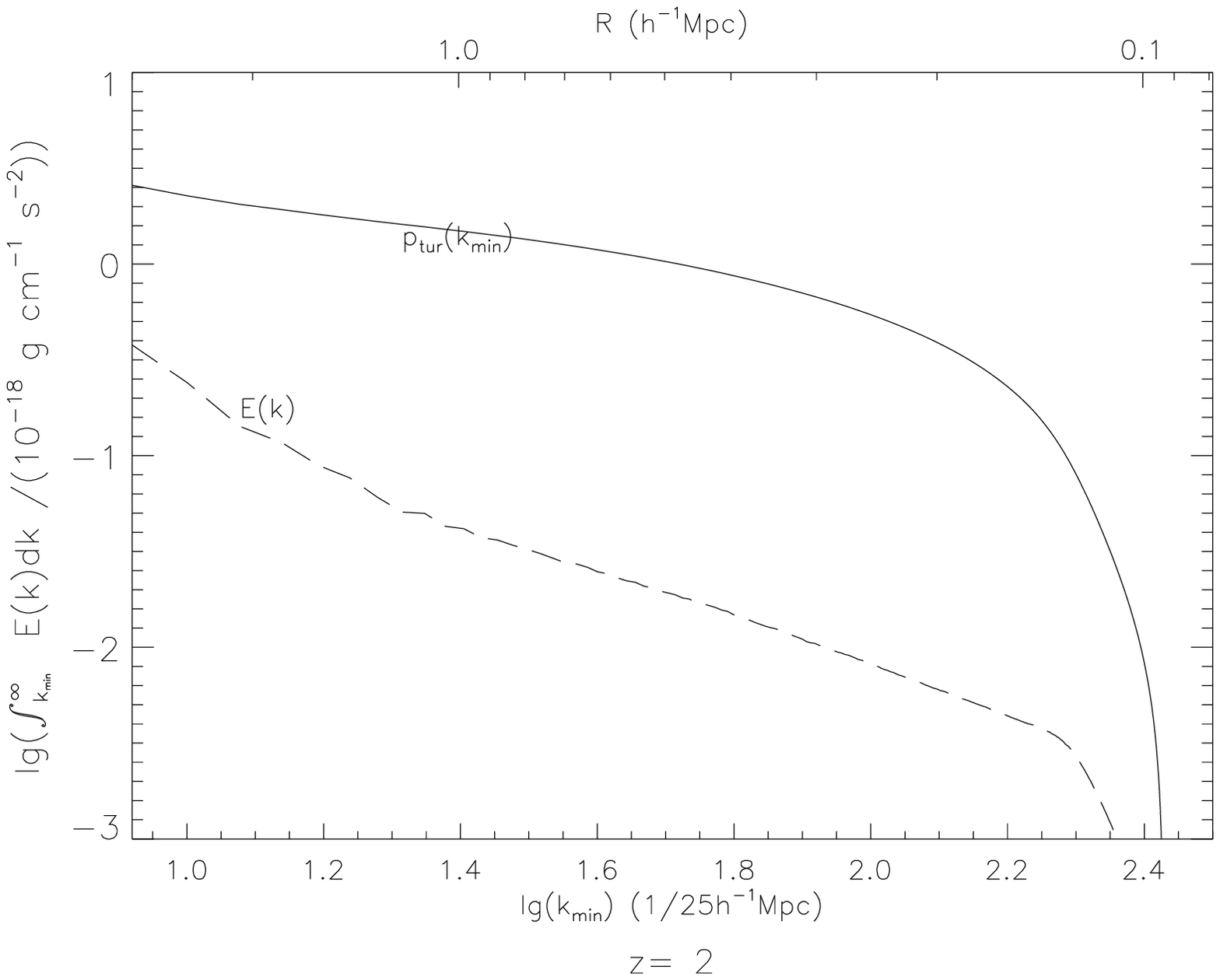}
\includegraphics[width=8cm]{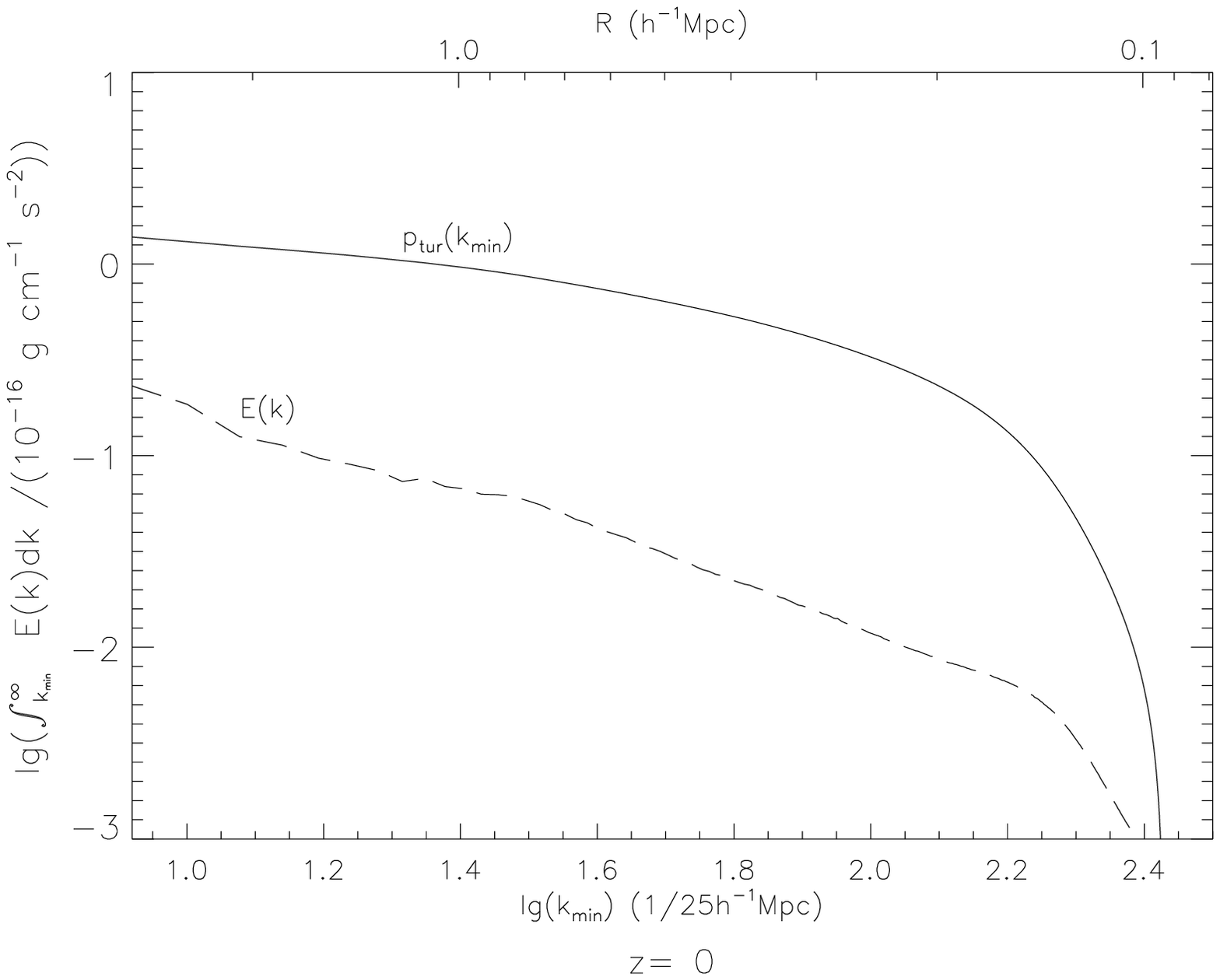}
\end{center}
\caption{The spectrum of turbulent pressure $p_{tur}(k_{min})$, given by 
$p_{tur}(k_{min})=\int_{k_{min}}^{k_{max}} E(k) dk$, at redshifts
$z=2$ (left) and $z=0$ (right). The energy spectra $E(k)$ are also shown 
in each panel.
}
\end{figure}

The scale-dependence of the turbulent pressure is very weak. A
decrease of one order of magnitude in scales from $R=3$ to $0.3$
$h^{-1}$ Mpc can only lead to deceases in the pressure $p_{tur}$ by
a factor of 4. On the other hand, the mass $m$ of a cluster is related 
to its scale radius $r_s$ approximately as $m \propto r_s^3$ (Cooray \& 
Sheth, 2002). The gravitational potential of $m$ halos at $r_{s}$ is 
$Gm/r_{s}\propto r_{s}^{2}$. Therefore, the ratio of the turbulent pressure 
to the gravitational potential at $r_s$ would be larger for clusters with 
smaller mass $m$. The effect of turbulent pressure on gravitational 
collapsing of baryon gas would be more significant on smaller clusters.

%fig9
\begin{figure}[htb]
\begin{center}
\includegraphics[width=8cm]{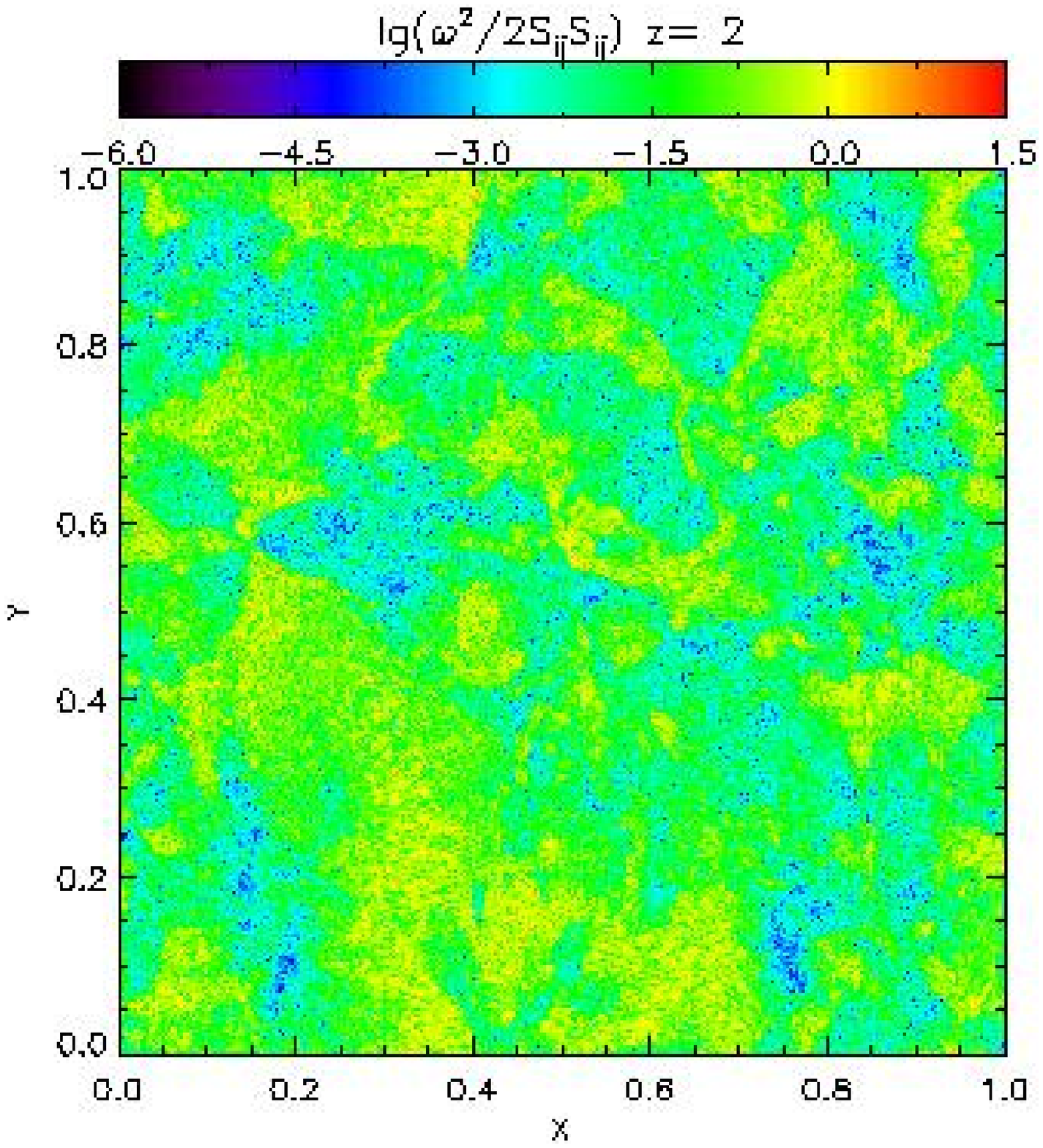}
\includegraphics[width=8cm]{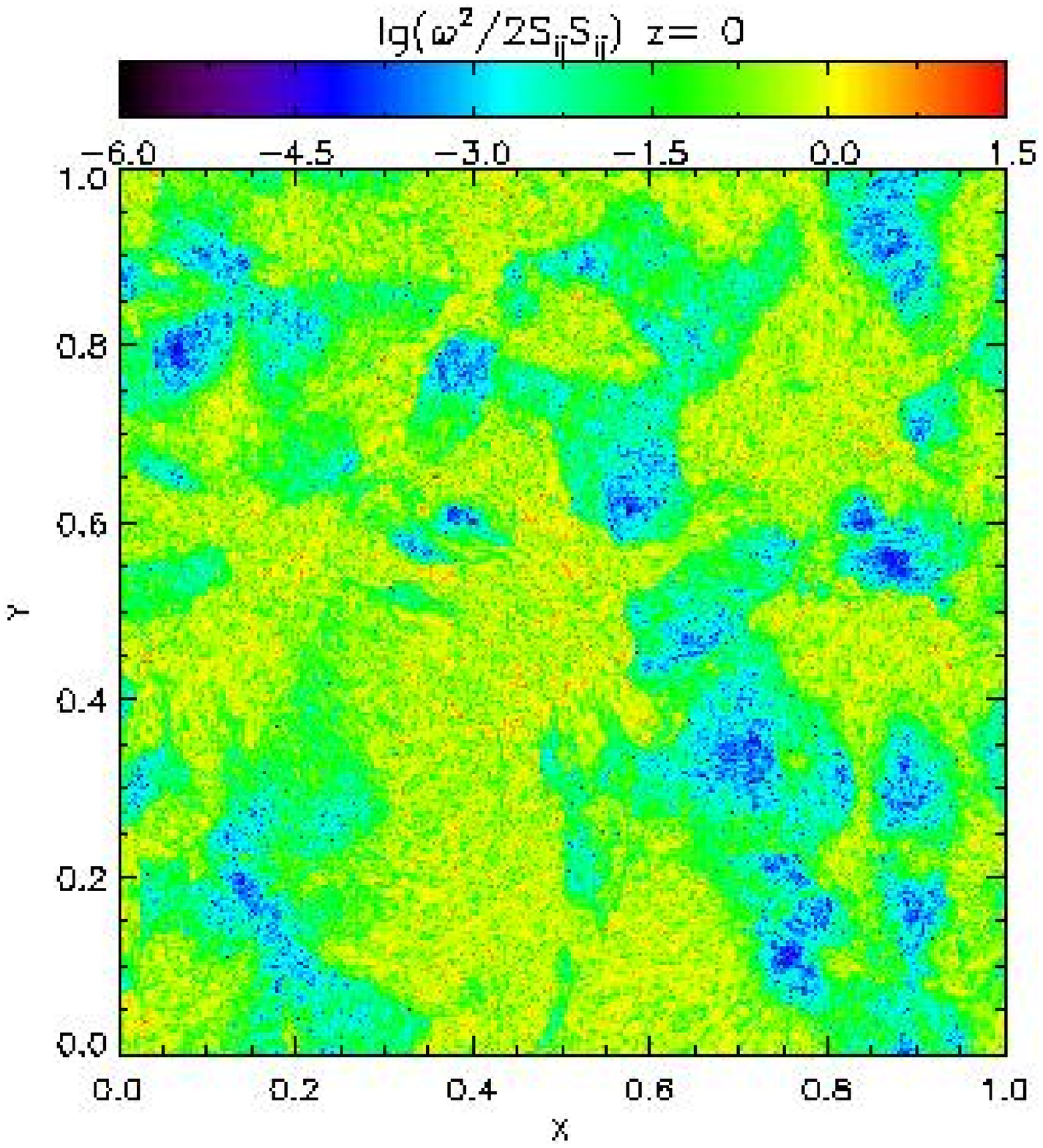}
\end{center}
\caption{The distribution of $\ln(\omega^2/2 S_{ij}S_{ij})$, which characterizes
the net effect of turbulence on clustering and positive value represents prevention
, in a 2-D slide of $25\times 25\times 0.1 h^{-3}$ Mpc$^3$ at redshift $z=2$ (left)
and $0$ (right).}
\end{figure}

\subsection{Vorticity and the Growth Rate of Velocity Divergence}

In the nonlinear regime of the IGM gravitational clustering, the dynamical effect of
turbulence can be estimated by eq.(3). Here, we are focusing on the first two terms
from vorticity and strain rate. As discussed in \S 2.2, the net effect on the
clustering is determined by the sign of quantity,
%eq
\begin{equation}
\frac{1}{2}\omega^2- S_{ij}S_{ij}=\frac {1}{2}
[(\partial_iv_j)(\partial_iv_j)-3(\partial_j v_i)(\partial_iv_j)].
\end{equation}
For a Gaussian velocity field, we have $\langle 3(\partial_j
v_i)(\partial_iv_j)\rangle=\langle
(\partial_iv_j)(\partial_iv_j)\rangle$, and in average, the net effect
of velocity field in eq.(3) is statistically null. However, for a non-Gaussian
velocity field, it can be either positive or
negative, which is dependent on the property of the velocity field.

In homogeneous and isotropic turbulence, as $\langle (\partial_j
v_i)(\partial_iv_j)\rangle =0$ (Batchelor 1959), the signs of
Eq.(14) are always positive, which results in prevention of
gravitational collapsing in the IGM.  Figure 9 plots the spatial
distribution of $\ln(\omega^2/2 S_{ij}S_{ij})$ in the same
simulation slide as that used in Figure 3. Comparing Figure 9 with
Figures 3 , we find that all those cells with $\ln(\omega^2/2
S_{ij}S_{ij})>0$ are located in the clouds around density peaks,
where the vorticity is dominant. It provides a mechanism to prevent
or slow down the IGM clustering with respect to the underlying dark
matter.

We search for cells with $(1/2)\omega^2- S_{ij}S_{ij}>0$. At redshift $z=0$,
there is a fraction $7.6\%$ of volume, $16.6\%$ of mass, with positive values of
$(1/2)\omega^2- S_{ij}S_{ij}$, while at redshift $z=2$, this volume fraction has decreased
down to $2.6\%$. Thus, the effect of turbulence becomes stronger to prevent
 the IGM clustering at lower redshifts.

%fig10
\begin{figure}[htb]
\begin{center}
\includegraphics[width=8cm,angle=-90]{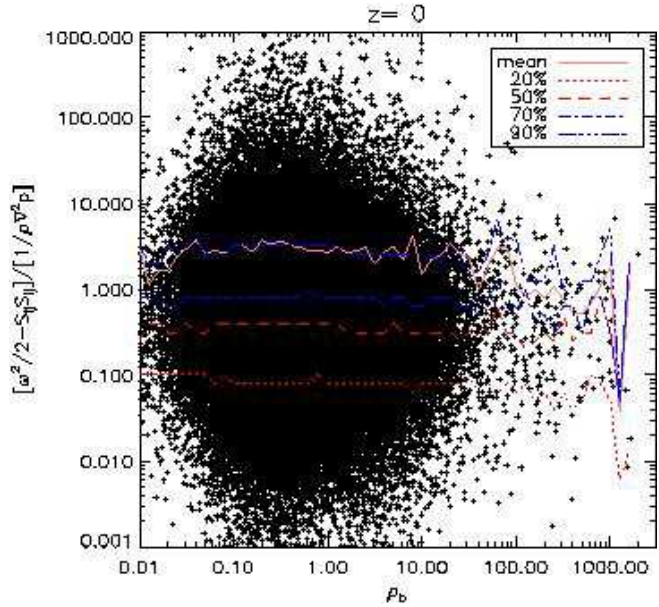}
\end{center}
\caption{Comparison of effects of turbulent pressure and thermal pressure
on the divergence $d$ (eq.(3)), $[\frac{1}{2}\omega^{2}-S_{ij}S_{ij}]/[\frac{1}{\rho}\nabla^{2}p]$, to
baryon density of randomly selected cells with $\frac{1}{2}\omega^{2} > S_{ij}S_{ij}$ 
at redshift $z=0$. Solid line gives the mean value of this ratio at 
every density bin. Broken lines give the cumulative probability $20\%$,$50\%$,$70\%$ and $90\%$, from
bottom to up.}
\end{figure}

In order to compare the effects of turbulent pressure and thermal pressure on the 
divergence $d$, we calculate the ratio of $(1/2)\omega^2- S_{ij}S_{ij}$ to $-\nabla^2 p$ 
, taken from eq.(3), cell by cell. The result is presented in Figure 10, in which the cells 
are randomly selected from those with $\frac{1}{2}\omega^{2} > S_{ij}S_{ij}$ in the whole
simulation samples. We find that for most densities nearly $30\%$ of those cells with  
$\frac{1}{2}\omega^{2} > S_{ij}S_{ij}$ have a value of this ratio larger than $1$ and indicate that 
mean turbulent pressure dominates over thermal pressure in them. Obviously, the dynamical 
prevention provided by turbulence  could be comparable to that of thermal pressure and even
become dominate in a considerable fraction of the whole volume.

\section{Discussions and Concluding Remarks}

The relationship between the fields of vorticity and velocity is
similar to the relationship between the current of charge
density and magnetic field, and thus, vorticity would be a
measurement of the coherent spatial structures of velocity field.
Moreover, the dynamical equation of vorticity is free from the
gravitational field of dark matter and cosmic fluid. These
remarkable features are very useful to study the nonlinear behavior
of cosmic baryon fluid, especially the clustering behavior of the
turbulent cosmic fluid in the gravitational field of underlying dark
matter.

We show that the vorticity field of baryonic
matter is significantly increasing with time when redshift $z \leq
2$. It can be understood that vorticity is effectively generated by shocks
and complex structures of the baryon fluid, and then amplified by
the rate-of-strain. At redshift $z=4$, the largest vorticity is only
of the order of $\omega t\simeq 10$, while it is $\omega t\simeq
10^2$ at present universe. The IGM vorticity field is
non-Gaussian and intermittent at all redshifts. The PDF of vorticity
evolves from approximately exponential distributions at high redshifts to
a distribution with log-normal long tail at present epoch.

The spatial configuration of the vorticity field is found to be very different
from that of velocity and mass density. The distribution of vorticity does not
follow the underlying matter structures, such as filaments and sheets. It
always shows 3-D cloudy structures around gravitational collapsed regions,
i.e. the knots in the filament-sheets structures. Even in regions surrounding
high density structures, vorticity can be strong because complex structures,
such as curved shocks and collision of shocks, are already formed around knots
at their early phase of formation. Vorticity would be more effective
to reveal the clustering behavior, which is overlooked by the mass density
field in some way.

The fluctuations of vorticity field is useful to measure the
development of a fully developed turbulence in the cosmic fluid. The
relation between the power spectra of vorticity and velocity
provides a measurement on the scale of velocity fluctuations where
turbulence is fully developed. We find that the cosmic fluid is in
the state of fully developed homogeneous and isotropic turbulence in
the scale range of $0.2 h^{-1}$Mpc to $3.0 h^{-1}$Mpc at present
epoch. With this result, we calculate the turbulent pressure. It is
of the order of $1.5\times 10^{-17}$ g cm$^{-1}$s$^{-2}$ at $z=0$ in
average, which is equivalent to the thermal pressure of gas with
mean cosmic baryon density at temperature $1.0\times 10^6$ K. It
tends to slow down the gravitational clustering of the baryon fluid.
Moreover, the spectrum of turbulence pressure is weakly dependent
on scale $k$, and then the effect of turbulent pressure would be relatively
stronger on smaller objects.

The turbulent pressure may shed light on the problem of
overcooling,  i.e. the fraction of cold gas and stars in
regions of galaxies, galaxies groups and clusters given by $\Lambda$CDM
simulations is significantly higher than the observed value at $z\sim 0$
( Nagai \& Kravtsov, 2004, Crain et al. 2007, Keres et al. 2009). A possible 
way to solve this problem is to assume that the IGM undergo a pre-heating at low
redshift (e.g. de Silva et al 2004). However, the pre-heating model is
strongly in contradiction with the observations of the low-redshift
Ly$\alpha$ forest of quasars, which cannot exist if the temperature
of the IGM is $\geq 10^5$ K. Galactic winds is another mechanism proposed
to suppress star formation in galaxies. Hydrodynamic simulations, however,
suggest that such feedback would be inefficient in galaxies with
$M_{gal} \geq 10^{9} M_{\odot}$(Mac Low \& Ferrara 1999).
Turbulent pressure essentially is dynamical and nonthermal. It can
play the similar role as thermal pressure to prevent the gravitational
clustering, while does not affect the thermal state and ionizing process
of hydrogen in the IGM. The turbulent IGM can remain a temperature of
$10^{4-5}$K and hence consistent with the observation of Ly$\alpha$ forest.
If the IGM is turbulent, the Ly$\alpha$ absorption lines will not only show
thermal broadening but also turbulent broadening. Observation of
Ly$\alpha$ line widths of HI and HeII indicates that the broadening
of Ly$\alpha$ forest is partially given turbulence broadening (Shull
et al 2004, Zheng et al 2004, Liu et al 2006).

Vorticity enhances the transportation of mass, momentum and kinetic
energy. The cascade of vortical structures leads to transfer of  kinetic energy
of vortical motion from large scales to small scales. The
turbulence energy will further dissipate into thermal motion.
This processes will enhance efficiently the entropy production via
the thermalization and virialization. In addition, the turbulent
motion can cause diffusive mixing of materials, which tends to wipe
out gradients in the distribution of chemical composition. The
details will be reported in the near future.

\begin{acknowledgements}

We thanks Dr. Priya Jamkhedkar for her helps. WSZ acknowledges the
support of the International Center for Relativistic Center Network
(ICRAnet). This work is partially supported by the National Science
Foundation of China grant  NSFC 10633040,  10725314 , 10621303 and
the  973 Program under contract No. 2007CB815402.

\end{acknowledgements}

\appendix

\section{The basic equations}

\subsection{Hydrodynamical equations for the IGM}

The IGM is assumed to be an ideal fluid with polytropic index
$\gamma$. The hydrodynamic equations for the IGM in the expanding
universe can be written in the following form
%eq1
\begin{equation}
\label{hydro} \dot{U}+\partial_i f_{i}[U]=f(t,U)
\end{equation}
where $\partial_i\equiv\partial/\partial X_{i}$ ($i=1,2,3$), $X_{i}$
denote the proper coordinates, which are related to comoving coordinates
$x_{i}$ by $X_{i}=a(t)x_{i}$, $a(t)$ being the scale factor. The quantity
$U$ in eq.(A1) contains five components as
%eq2
\begin{equation}
U=(\rho, \rho{\bf v}, E)
\end{equation}
where $\rho$ is the comoving density of the IGM, ${\bf v}=\{v_i\}$
($i=1,2,3$)  are the peculiar velocity on three axes,
$E=P/(\gamma-1)+\frac{1}{2}\rho{\bf v}^2$ is the total energy per
unit comoving volume, $P=a^3p$, and $p$ is the pressure of the IGM.
The quantities $f_{i}(U)$ in Eq.(A1) are given by the conservation
laws of mass, momentum and energy as
%eq3
\begin{eqnarray}
f_{1}(U)&=&[\rho v_{1}, \rho (v_{1})^2+P, \rho v_{1}v_{2}, \rho v_{1}v_{3}, v_{1}(E+P)]
   \nonumber \\
f_{2}(U)&=&[\rho v_{2}, \rho v_{1}v_{2}, \rho (v_{2})^2+P, \rho v_{2}v_{3}, v_{2}(E+P)]
  \nonumber \\
f_{3}(U)&=&[\rho v_{3}, \rho v_{1}v_{3}, \rho v_{2}v_{3}, \rho (v_{3})^2+P, v_{3}(E+P)]
\end{eqnarray}
The  ''force" term $f(t, U)$ on the right hand side
of Eq. (A1) is given by
%eq4
\begin{equation}
f(t,U)=(0,-\frac{\dot a}{a}\rho{\bf v}+\frac{1}{a}\rho{\bf g}, -2\frac{\dot
a}{a}E+\frac{1}{a}\rho{\bf v}\cdot{\bf g}-\Lambda_{rad}).
\end{equation}
The term of $-(\dot a/a)\rho{\bf v}$ is from the expansion of
the universe. The term of $\Lambda_{rad}$ in Eq.(A4) is given by
the radiative heating-cooling of the baryon gas. The gravitational
force ${\bf g}=-\nabla \phi$ is produced by the matter including CDM and baryon
, given by
%eq5
\begin{equation}
\nabla^2 \phi = \frac{4\pi G}{a} \bar{\rho}_{tot}\delta_{tot}.
\end{equation}
where the operator $\nabla$ acts on the comoving coordinate ${\bf x}$.
$\delta_{tot}=[\rho_{tot}({\bf x},t)-\bar{\rho}_{tot}]/\bar{\rho}_{tot}$,
and $\rho_{tot}$ is the total comoving mass
density. Its mean value is
$\bar{\rho}_{tot}(t) =1/6\pi Gt^2 \propto a^{-3}$. The gravitational
potential $\phi$ is zero (or constant) when the density
perturbation $\delta_{tot}=0$.

\subsection{Vorticity equation}

Euler equation in comoving coordinates
%eq1
\begin{equation}
 \partial_t {\rho v_{i}}+\frac{1}{a}(\partial_j \rho v_{j}v_{i}+\partial_j\delta_{ij}P)
=-\frac{\dot a}{a}\rho v_{i}+\frac{1}{a}\rho g_{i},
\end{equation}
or
%eq2
\begin{equation}
 \rho\partial_t { v_{i}}+  v_{i}\partial_t\rho+
 \frac{1}{a}(v_{i}\partial_j \rho v_{i}+\rho v^j\partial_jv_{i}+ \partial_j\delta_{ij}P)
=-\frac{\dot a}{a}\rho v_{i}+\frac{1}{a}\rho
 g_{i},
\end{equation}
,where $\partial_t\equiv\partial/\partial t$.
Therefore
\begin{equation}
 \rho\partial_t { v_{i}}+\frac{1}{a}(\rho v_{j}\partial_jv_{i}+ \partial_j\delta_{ij}P)
=-\frac{\dot a}{a}\rho
v_{i}+\frac{1}{a}\rho
 g_{i},
\end{equation}
or
\begin{equation}
%\partial_t { v^i}+\frac{1}{a}v^j\partial_jv^i= -\frac{1}{a}\frac{1}{\rho}
\partial_j\delta_{ij}P
%-\frac{\dot a}{a} v^i+ g^i.
\partial_t { v_{i}}+\frac{1}{a}v_{j}\partial_jv_{i}= -\frac{1}{a}(\frac{1}
{\rho}\partial_j\delta_{ij}P
+\dot a v_{i}- g_{i}).
\end{equation}

Using Levi Civita symbol
\begin{equation}
\epsilon_{ijk}\epsilon_{ilm}=\delta_{jl}\delta_{km}-\delta_{jm}\delta_{kl}
\end{equation}
we have
\begin{equation}
v_{j}\partial_jv_{i}=\frac{1}{2}\partial_iv_{j}v_{j} -
\epsilon_{ijk}v_j\omega_k,
\end{equation}
where $\omega_{i}=\epsilon_{ijk}\partial^jv_{k}$ is vorticity.

Taking operator of curl $\epsilon_{ijk}\partial^j$ on eq.(A9), we
have term by term.
\begin{equation}
\epsilon_{ijk}\partial^j\partial_t {
v_{k}}=\partial_t\epsilon_{ijk}\partial^jv_{k}=\partial_t\omega_{i}
\end{equation}
\begin{equation}
\epsilon_{ijk}\partial^jv_{l}\partial_lv_{k}=\epsilon_{ijk}\partial^j
\frac{1}{2}\partial_kv_{l}v_{l} -
\epsilon_{ijk}\partial^j\epsilon_{klm}v_l\omega_m
\end{equation}
\begin{equation}
\epsilon_{ijk}\partial^j \frac{1}{2}\partial_kv_{l}v_{l}=0
\end{equation}
\begin{equation}
\epsilon_{ijk}\partial^j\epsilon_{klm}v_l\omega_m=
\epsilon_{kij}\epsilon_{klm}\partial^jv_l\omega_m
=\partial^mv_i\omega_m-\partial^lv_l\omega_i=\omega_m\partial^m
v_i-\omega_i\partial^lv_l-v_l\partial^l \omega_i
\end{equation}

Therefore, we have vorticity equation as
\begin{equation}
\partial_t\omega_i +\frac{1}{a}v_l\partial^l \omega_i=\frac{1}{a}(\omega_m\partial^m
v_i-\omega_i\partial^lv_l+\frac{1}{\rho^2}\epsilon_{ijk}\partial_j\rho\partial_k
p- \dot{a}\omega_i).
%p- \frac{\dot{a}}{a}\omega_i.
\end{equation}
Because $\omega_i(\partial^j v_i-\partial^i v_j)=0$, we have
\begin{equation}
\partial_t\omega_i +\frac{1}{a}v_l\partial^l \omega_i=\frac{1}{a}(S_{ij}
\omega_j-d \omega_i+
\frac{1}{\rho^2}\epsilon_{ijk}\partial_j\rho\partial_k
p- \dot{a}\omega_i)
%p- \frac{\dot{a}}{a}\omega_i
\end{equation}
where
\begin{equation}
S_{ij}=\frac{1}{2}(\partial^j v_i+\partial^i v_j)
\end{equation}
and $d=\partial_iv_{i}$. In vector format
\begin{equation}
\partial_t{\bf \omega} +\frac{1}{a}{\bf v}\cdot{\nabla}\omega=\frac{1}{a}({\bf S}
\cdot {\bf \omega}
-d {\bf \omega}
+\frac{1}{\rho^2}\nabla \rho \times\nabla p-\dot{a}\omega)
%+\frac{1}{\rho^2}\nabla \rho \times\nabla p-\frac{\dot{a}}{a}\omega)
\end{equation}

\subsection{Equation of divergence}

Taking operator $\partial^i$ on eq.(A9), we have term by term,
\begin{equation}
\partial^i\partial_t {
v_{i}}=\partial_t d
\end{equation}
\begin{equation}
\partial^iv_{j}\partial_jv_{i}=v_{j}\partial_j\partial^iv_{i}+(\partial^iv_{j})(\partial_jv_{i})
=v_{j}\partial_jd+(\partial^iv_{j})(\partial_jv_{i})
\end{equation}
Using
\begin{equation}
\partial^i v_j=\frac{1}{2}(\partial^i v_j+\partial^j v_i)+\frac{1}{2}(\partial^i
v_j-\partial^jv_i),
\end{equation}
we have
\begin{equation}
(\partial^iv_{j})(\partial_jv_{i})=S_{ij}S_{ij}+\frac{1}{4}(\partial^i
v_j-\partial^jv_i)(\partial^j v_i-\partial^i v_j)=S_{ij}S_{ij}+
\frac{1}{2}\epsilon_{ijk}\partial_jv_k\epsilon_{ilm}\partial_lv_m.
\end{equation}
Therefore, the equation of divergence is

\begin{equation}
\partial_td +\frac{1}{a}v_l\partial^l d=\frac{1}{a}(\frac{1}{2}\omega_i
\omega_i-S_{ij}S_{ij}-
\frac{1}{\rho}\partial_i\partial_i p
+\frac{1}{\rho^2}\partial_j\rho\partial_j p-\dot{a}d
-\frac{4\pi G}{a}(\rho-\rho_0)).
\end{equation}
or in vector format
\begin{equation}
\partial_td +\frac{1}{a}{\bf v}\cdot{\bf \nabla} d=\frac{1}{a}(\frac{1}{2}
{\bf \omega}\cdot{\bf \omega}-
S_{ij}S_{ij}-\frac{1}{\rho}\nabla^2 p
+\frac{1}{\rho^2}(\nabla\rho)\cdot(\nabla p)-\dot{a}d
-\frac{4\pi G}{a}(\rho-\rho_0)).
\end{equation}

\subsection{A brief description of the numerical algorithm.}

We use the fifth order finite difference WENO scheme (Jiang \& Shu 2006) to
demonstrate the basic idea of the WENO methodology. The fifth order
WENO finite difference spatial discretization to a conservation law
such as
%eqA26
\begin{equation}
u_t + f(u)_x + g(u)_y + h(u)_z = 0
\end{equation}
approximates the derivatives, for example $f(u)_x$, by a
conservative difference

%eqA27

\begin{equation}
f(u)_x |_{x=x_j} \approx \frac{1}{\Delta x} \left( \hat{f}_{j+1/2} -
\hat{f}_{j-1/2} \right)
\end{equation}

along the $x$ axis, with $y$ and $z$ fixed, where $\hat{f}_{j+1/2}$
is the numerical flux. $g(u)_y$ and $h(u)_z$ are approximated in the
same way.  Hence finite difference methods have the same format for
one and several space dimensions, which is a major advantage.  For
the simplest case of a scalar equation (A26) and if $f'(u)
\geq 0$, the fifth order finite difference WENO scheme has the flux
given by
\begin{equation}
\hat{f}_{j+1/2} = w_1 \hat{f}_{j+1/2}^{(1)} + w_2
\hat{f}_{j+1/2}^{(2)} + w_3 \hat{f}_{j+1/2}^{(3)}
\end{equation}
where $\hat{f}_{j+1/2}^{(i)}$ are three third order accurate fluxes
on three different stencils given by
\begin{eqnarray}
\hat{f}_{j+1/2}^{(1)}  = \frac{1}{3} f(u_{j-2}) - \frac{7}{6}
f(u_{j-1}) + \frac{11}{6} f(u_{j}), \\
\hat{f}_{j+1/2}^{(2)}  = -\frac{1}{6} f(u_{j-1}) + \frac{5}{6}
f(u_{j}) + \frac{1}{3} f(u_{j+1}),\\
\hat{f}_{j+1/2}^{(3)}  = \frac{1}{3} f(u_{j}) + \frac{5}{6}
f(u_{j+1}) - \frac{1}{6} f(u_{j+2}).
\end{eqnarray}
Notice that the combined stencil for the flux $\hat{f}_{j+1/2}$ is
biased to the left, which is upwinding for the positive wind
direction due to the assumption $f'(u) \geq 0$. The key ingredient
for the success of WENO scheme relies on the design of the nonlinear
weights $w_i$, which are given by
\begin{equation}
w_i = \frac {\tilde{w}_i}{\sum_{k=1}^3 \tilde{w}_k},\qquad
 \tilde{w}_k = \frac {\gamma_k}{(\varepsilon + \beta_k)^2} ,
\end{equation}
where the linear weights $\gamma_k$ are chosen to yield fifth order
accuracy when combining three third order accurate fluxes, and are
given by
\begin{equation}
\gamma_1=\frac{1}{10}, \qquad \gamma_2=\frac{3}{5}, \qquad
\gamma_3=\frac{3}{10} ;
\end{equation}
the smoothness indicators $\beta_k$ are given by
\begin{eqnarray}
\beta_1 & = & \frac{13}{12} \left( f(u_{j-2}) - 2 f(u_{j-1})
+ f(u_{j}) \right)^2 +\frac{1}{4} \left( f(u_{j-2}) - 4 f(u_{j-1})
+ 3 f(u_{j}) \right)^2  \\
\beta_2 & = & \frac{13}{12} \left( f(u_{j-1}) - 2 f(u_{j})
+ f(u_{j+1}) \right)^2 +\frac{1}{4} \left( f(u_{j-1})
-  f(u_{j+1}) \right)^2  \\
\beta_3 & = & \frac{13}{12} \left( f(u_{j}) - 2 f(u_{j+1})
+ f(u_{j+2}) \right)^2 +\frac{1}{4} \left( 3 f(u_{j}) - 4 f(u_{j+1})
+ f(u_{j+2}) \right)^2 ,
\end{eqnarray}
and they measure how smooth the approximation based on a specific
stencil is in the target cell. Finally, $\varepsilon$ is a parameter
to avoid the denominator to become 0 and is usually taken as
$\varepsilon = 10^{-6}$ in the computation.  There are no other
parameters needed to be tuned by the user in the WENO method.

Meanwhile, the time step in simulation is set to the minimum value among
two time scales. One is given by Courant condition as
\begin{equation}
\Delta t_{cfl} \leq \frac{CFL[a(t)\Delta x ]}{max(|v_{1}+c_{s},v_{2}+c_{s},v_{3}+c_{s})},
\end{equation}
where $\Delta x$ is the cell size, $c_{s}$ is the local sound speed,
$v_{1}$, $v_{2}$, and $v_{3}$ are fluid velocities, and CFL is the
Courant number, here $CFL=0.60$. The other one is from cosmic
expansion, which requires that $\Delta a/a < 0.02$ within a single
time step.

The WENO scheme is proven to be uniformly
fifth order accurate including at smooth extrema, and this is
verified numerically. Near discontinuities the scheme produces sharp
and non-oscillatory discontinuity transition. The approximation is
self-similar. Namely, when fully discretized with the Runge-Kutta
methods, the scheme is invariant when the spatial
and time variables are scaled by the same factor. This is a major
advantage for approximating conservation laws which are invariant
under such scaling.

\end{document}